\documentclass{jfm}

\usepackage{graphicx}
\usepackage{natbib}
\usepackage{graphics} 
\usepackage{epsfig} 
\usepackage{amsmath} 
\usepackage{amssymb}  
\usepackage{amsbsy}
\usepackage{braket}
\usepackage{array}
\usepackage{color}
\usepackage{balance}
\usepackage{setspace}
\usepackage{enumerate}
\usepackage{upgreek}
\newcommand{\pderiv}[2]{\frac{\partial #1}{\partial #2}}
\newcommand{\norm}[1]{\left\Vert#1\right\Vert}

\ifCUPmtlplainloaded \else
  \checkfont{eurm10}
  \iffontfound
    \IfFileExists{upmath.sty}
      {\typeout{^^JFound AMS Euler Roman fonts on the system,
                   using the 'upmath' package.^^J}%
       \usepackage{upmath}}
      {\typeout{^^JFound AMS Euler Roman fonts on the system, but you
                   dont seem to have the}%
       \typeout{'upmath' package installed. JFM.cls can take advantage
                 of these fonts,^^Jif you use 'upmath' package.^^J}%
      }
  \else
  \fi
\fi

\ifCUPmtlplainloaded \else
  \checkfont{msam10}
  \iffontfound
    \IfFileExists{amssymb.sty}
      {\typeout{^^JFound AMS Symbol fonts on the system, using the
                'amssymb' package.^^J}%
       \usepackage{amssymb}%
         \let\leq=\leqslant
         \let\geq=\geqslant
      }{}
  \fi
\fi

\ifCUPmtlplainloaded \else
  \IfFileExists{amsbsy.sty}
    {\typeout{^^JFound the 'amsbsy' package on the system, using it.^^J}%
     \usepackage{amsbsy}}
    {\providecommand\boldsymbol[1]{\mbox{\boldmath $##1$}}}
\fi

\providecommand\bnabla{\boldsymbol{\nabla}}
\providecommand\bcdot{\boldsymbol{\cdot}}

\newcommand\Rey{\mbox{\textit{Re}}}  

\newsavebox{\astrutbox}
\sbox{\astrutbox}{\rule[-5pt]{0pt}{20pt}}

\title{Modelling for Robust Feedback Control of Fluid Flows}

\author[B.~Ll.~Jones, P.~H.~Heins, E.~C.~Kerrigan, J.~F.~Morrison and~A.~S.~Sharma]%
{B\ls R\ls Y\ls N\ns Ll.\ns J\ls O\ls N\ls E\ls S$^1$%
  \thanks{Email address for correspondence: b.l.jones@sheffield.ac.uk},\ns
P.\ns H.\ns H\ls E\ls I\ls N\ls S$^1$,\ns
E.\ns C.\ns K\ls E\ls R\ls R\ls I\ls G\ls A\ls N$^{2,3}$\break
J.\ns F.\ns M\ls O\ls R\ls R\ls I\ls S\ls O\ls N$^3$\ns \and A.\ns S.\ns S\ls H\ls A\ls R\ls M\ls A$^4$}

\affiliation{$^1$Department of Automatic Control and Systems Engineering, University of Sheffield, Sheffield, S1 3JD, UK\\[\affilskip]
$^2$ Department of Electrical and Electronic Engineering, Imperial College London, SW7 2AZ, UK\\[\affilskip]
$3$ Department of Aeronautics, Imperial College London, London, SW7 2AZ, UK\\[\affilskip]
$4$ Engineering and the Environment, University of Southampton, Highfield, Southampton, SO17 1BJ, UK}

\pubyear{2014}
\volume{650}
\pagerange{119--126}
\date{?; revised ?; accepted ?. - To be entered by editorial office}
\begin{document}

\maketitle

\begin{abstract}
This paper addresses the problem of designing low-order and linear robust feedback controllers that provide~\emph{a priori} guarantees with respect to stability and performance when applied to a fluid flow. This is challenging since whilst many flows are governed by a set of nonlinear, partial differential-algebraic equations (the Navier-Stokes equations), the majority of established control system design assumes models of much greater simplicity, in that they are firstly: linear, secondly: described by ordinary differential equations, and thirdly: finite-dimensional. With this in mind, we present a set of techniques that enables the disparity between such models and the underlying flow system to be quantified in a fashion that informs the subsequent design of feedback flow controllers, specifically those based on the~$\mathcal{H}_\infty$ loop-shaping approach. Highlights include the application of a model refinement technique as a means of obtaining low-order models with an associated bound that quantifies the closed-loop degradation incurred by using such finite-dimensional approximations of the underlying flow. In addition, we demonstrate how the influence of the nonlinearity of the flow can be attenuated by a linear feedback controller that employs high loop gain over a select frequency range, and offer an explanation for this in terms of Landahl's theory of sheared turbulence. To illustrate the application of these techniques, a~$\mathcal{H}_\infty$ loop-shaping controller is designed and applied to the problem of reducing perturbation wall-shear stress in plane channel flow. DNS results demonstrate robust attenuation of the perturbation shear-stresses across a wide range of Reynolds numbers with a single, linear controller.
\end{abstract}


\section{Introduction}\label{Sec1}
The ability to exert control over fluid flows has received renewed attention in recent years, with the potential to improve the efficiency of fluid-based systems thereby offering wide-ranging economic and environmental benefits across a range of industries. Examples include the lowering of fuel costs and greenhouse gas emissions via the drag reduction of aircraft~\citep{Bushnell03} and shipping~\citep{Corbett03}, optimal mixing of chemical reagents~\citep{Couchman10} and wind turbine gust alleviation~\citep{Frederick10}, with many more examples stemming from the natural world~\citep{Fish06}. Attempts to control fluid flow are typically classified into three broad categories \citep{gadelhak}: passive~(e.g.~\citet{Choi93}), active open-loop~(e.g.~\citet{Sturzebecher03, Hanson10}) and active closed-loop control~(e.g.~\citet{Bewley2001, Hogberg, kim03, Kim07, Semeraro11}), each with their own merits and extensively discussed in many review papers and textbooks~(e.g.~\citet{Bewley2001, Collis2004, gadelhak}).

This paper is concerned with the use of active (in the sense that powered actuators are assumed) closed-loop control of fluid flows. There are compelling reasons for employing such control, despite it being the most difficult to implement practically, owing to the dual requirements of sensing and actuation. Principal amongst these reasons is the unique ability of feedback controllers to reject the effects of~\emph{uncertainty} upon the desired outputs of a system~\citep{UncandFeed}, a concept that is of central importance in obtaining suitable control models for fluid flows, and which is the primary focus of this paper.

Uncertainties arise not only from the intrinsic model assumptions but also from exogenous disturbances inherent to practical problems.  To synthesise a feedback controller for a fluid flow, a model describing the dynamics of the system is required, where the system (or ``plant'') comprises actuators, sensors and the flow itself, in addition to the spatial interconnections between these subsystems. The dynamics of electromechanical components, such as pressure sensors~\citep{Arthur06} and synthetic jet actuators~\citep{Gallas03}, are typically well approximated by lumped-parameter models consisting of a few ordinary differential equations (ODEs). However, this is seldom the case for fluid flows, described in many cases by the incompressible Navier-Stokes equations: 
\begin{subequations}\label{navstokes}
\begin{align}
	\pderiv{\boldsymbol{V}(\boldsymbol{x},t)}{t}&=-\boldsymbol{V}(\boldsymbol{x},t)\bcdot\bnabla\boldsymbol{V}(\boldsymbol{x},t)-\nabla P(\boldsymbol{x},t)+\frac{1}{\Rey}\bnabla^2\boldsymbol{V}(\boldsymbol{x},t)+\boldsymbol{g}(\boldsymbol{x},t),\label{navstokesa}\\
	0&=\bnabla\bcdot \boldsymbol{V}(\boldsymbol{x},t),\label{navstokesb}
\end{align}
\end{subequations}
where~$\boldsymbol{V}(\boldsymbol{x},t)$ and~$P(\boldsymbol{x},t)$ are the velocity and pressure fields, respectively, evolving in domain~$\Omega\in\mathbb{R}^3$ under the influence of an external forcing~$\boldsymbol{g}(\boldsymbol{x},t)$, with~$\boldsymbol{x}\in\Omega$ and~$t\in\mathbb{R}_+$. Boundary and initial conditions are given as:
\begin{equation*}
	\boldsymbol{V}(\boldsymbol{x},t)=\boldsymbol{V}_\partial(\boldsymbol{x},t)~\mathrm{with}~\boldsymbol{x}\in\partial\Omega,\qquad\boldsymbol{V}(\boldsymbol{x},0)=\boldsymbol{V}_0,
\end{equation*}
where~$\partial\Omega$ is the boundary of the domain. In contrast to~\eqref{navstokes}, the majority of existing modern control systems theory relies upon models in standard, linear~\emph{state-space} form:
\begin{subequations}\label{standard}
\begin{align}
	\dot{\mathsfbi{x}}(t)&=\mathsfbi{A} \mathsfbi{x}(t)+\mathsfbi{B}\mathsfbi{u}(t),\label{standarda}\\
	\mathsfbi{y}(t)&=\mathsfbi{C}\mathsfbi{x}(t)+\mathsfbi{D}\mathsfbi{u}(t),\label{standardb}
\end{align}
\end{subequations}
where $\mathsfbi{A}\in\mathbb{R}^{n\times n}$, $\mathsfbi{B}\in\mathbb{R}^{n\times m}$, $\mathsfbi{C}\in\mathbb{R}^{q\times n}$, $\mathsfbi{D}\in\mathbb{R}^{q\times m}$, $\mathsfbi{x}(t)\in\mathbb{R}^n$ is the state vector with initial state $\mathsfbi{x}(0)=\mathsfbi{x}_0$, $\mathsfbi{u}(t)\in\mathbb{R}^m$ is the vector of control inputs and $\mathsfbi{y}(t)\in\mathbb{R}^q$ is the measurement vector. The states in~\eqref{standarda} evolve according to a finite-dimensional set of linear ODEs, and for the purposes of practical controller implementation it is desirable that the number of states be small, typically no more than~$n\sim \mathcal{O}(10^2)$. This means that in order to apply standard controller synthesis algorithms, the control model~\eqref{standard} must be of much greater simplicity than the underlying flow model~\eqref{navstokes}.

Attempts to approximate the plant~\eqref{navstokes} by the control model~\eqref{standard} represents a trade-off between reduced complexity for increased plant/model uncertainty. The process by which this is achieved gives rise to a further challenge, that is, the details of the transformation process itself. Although the ability to reduce the effects of uncertainty is an inherent feature of any control system employing feedback, certain branches of control theory handle the effects of uncertainty in a more rigorous fashion than others. Robust control~\citep{ZDG96,Zhou,Dullerud}, comprising a family of~$\mathcal{H}_\infty$ design methods~\citep{Glad,Skog} are of particular importance in this respect, and successful application of these methods has been demonstrated upon fluid flows~\citep{Bewley98,Baramov04,Lauga04,Bobba}. An attractive feature of robust control is its ability to provide a priori guarantees concerning the degree of stability of the closed-loop system, subject to model uncertainty and exogenous disturbances. The starting point for generating such controllers is a model of the form~\eqref{standard} that describes the linear dynamics of the flow.

\subsection{The importance of linear dynamics}\label{Sec1a}
A question that naturally arises is under what circumstances can a linear feedback controller, synthesised from a linear model~\eqref{standard}, actually stabilise a flow governed by~\eqref{navstokes}?  Although linearisations of~\eqref{navstokes} are inevitably unable to capture the nonlinear dynamics that endow turbulent flows with their `multiscale' characteristics~\citep{Kim07}, they are widely accepted as being relevant in explaining such phenomena as transition to turbulence in wall-bounded flows~\citep{Semeraro11,Butler, Trefethen, Schmid_Hen}, as well as at least some of the mechanisms that sustain turbulence in such flows.
In this respect, linear effects have received some attention since Batchelor and Proudman's seminal work on rapid distortion theory (RDT)~\citep{HC90,LeeKM90}. \citet{Farrell93, Farrell96} have suggested that the linearised Navier-Stokes equations in plane channel flow under stochastic forcing can exhibit behaviour reminiscent of the streamwise vortices and streaks characteristic of turbulent flow. Transient growth studies have highlighted the importance of the linear operator to streak formation \citep{Butler, Chernyshenko05}. The input-output (gain-based) analysis by \citet{Jovanovic05} of the linearised Navier-Stokes equations also revealed the importance of long streaky structures. ~\citet{Kim00} demonstrated in simulations of turbulent channel flow that the turbulence decays without the term coupling the wall-normal vorticity and the wall-normal velocity in the linearised Navier-Stokes equations.

\begin{figure}
 \centerline{\includegraphics[scale=0.425]{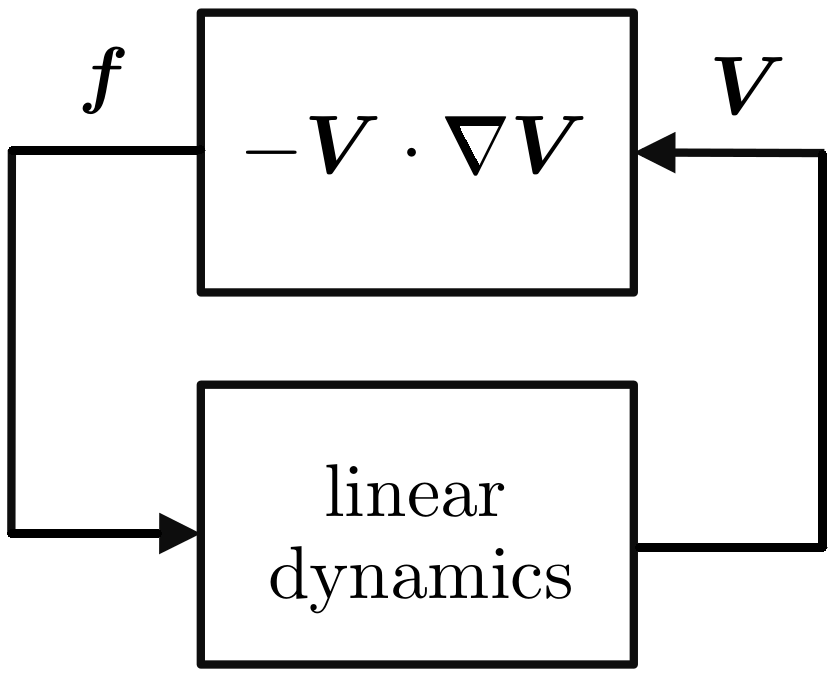}}
  \caption{System-level description of the turbulence process. The nonlinearity produces a force~$\boldsymbol{f}:=-\boldsymbol{V}\bcdot\bnabla\boldsymbol{V}$ that acts as a disturbance input to the linear subsystem.}
 \label{Fig1}
\end{figure}

More recently~\citet{McKeon10,McKeon12} explained how the structure of turbulence and its sustainment arises from a feedback interconnection between the linear and nonlinear terms of~\eqref{navstokesa}. This is depicted in~Figure~\ref{Fig1}, wherein the energy conserving nonlinearity forces a linear subsystem that describes the dynamics of fluctuations around a mean velocity profile~\citep{Sharma12}, with the linear dynamics playing a key role in selectively amplifying certain structures in the flow. In their analysis,~\citet{McKeon10} treated the nonlinearity of the flow arising from the interaction between scales as an unstructured forcing that acts to produce a turbulent mean profile of the appropriate form. By studying the singular vectors of the resolvent operator relating this input forcing to an output velocity field, these authors were able to predict coherent structures within the turbulent flow under study, that were in good agreement with experimental observations. The same authors also argued that the decomposition of the~Navier-Stokes equations into a linear system driven by an unknown forcing was justifiable owing to high gain at the critical layer resulting in the linear system being selective to the point where the exact form of the forcing was unimportant.
\citet{Sharma11} have also proved, using the passivity theorem, that a linear feedback controller can always be found to relaminarise a turbulent flow, given sufficient actuation and sensing. This was explained in physical terms with respect to~Landahl's theory of sheared turbulence~\citep{landahl77:pof,landahl75:siam,landahl67:jfm}.   Since the interaction of shear fluctuations is linear and underpins the generation of turbulent fluctuations, the linear control strategy is effective.  Shear interaction is a linear RDT approximation embodied in the Orr-Sommerfeld-Squire~(OSS) equations: it is governed by the wall-normal disturbance velocity which appears in the coupling term and which is related to the pressure via the linear `fast' source term in the Poisson equation for pressure fluctuations~\citep{kim89,DM03}. As a result, the response to forcing of the wall-normal velocity and pressure is rather quicker than that of both the streamwise or spanwise velocities. This occurs because the shear interaction timescale is considerably shorter than either the viscous or turbulence timescales~\citep{landahl77:pof} so that the Reynolds stresses are less effective.~\citet{BT56} have shown that pressure-gradient fluctuations drive the momentum field, appearing as spikes in the instantaneous mean-square acceleration.~\citet{Sharma11} have also shown that first, these fluctuations reach a maximum at $y^+\approx 20$, and second, the forcing is at a maximum at the same location.  This explains why the linear controller is effective even though it is operating on the wall-normal component alone. Landahl's theory~\citep{landahl75:siam,landahl67:jfm} also provides a `wave-guide' model of the viscous sublayer in which the least dispersive components are those of the wall-normal velocity component and pressure fields.  Clearly, understanding these linear mechanisms and the extent to which they are local to the wall has a significant bearing on potential drag-reduction strategies: for active, linear control, a fundamental appreciation of the shear-interaction timescale is a prerequisite and clearly, pressure is a key component to the interaction between the inner, wall region and the outer layer~\citep{T61,B67b,morrison:philtrans}.

Given the importance of suppressing turbulence for reducing skin-friction drag, much attention has been focussed on designing controllers for wall-bounded flows, particularly plane channel-flows~\citep{Bewley98,Lee01,Hogberg,Baramov04,Hoepffner,Kim07}. 
\citet{kim03} examined different types of Linear Quadratic Regulator (LQR), also for turbulent channel flow, to minimise (1) wall-shear stress fluctuations, (2) turbulent kinetic energy, and (3), the linear coupling term.  All resulted in significant drag reduction, a common feature being a weakening of quasi-streamwise vortices resulting in reduced high skin-friction extrema at the wall. There are many models of the near-wall cycle (see, for example,~\citet{hamilton06}) but all suggest that transient energy growth, as described by the OSS equations, provides a linear paradigm of near-wall turbulence~\citep{Butler}.
Central to our approach is that, for a model-based feedback controller to be successful, the role of linear dynamics can be exploited.  A key challenge is that much of our knowledge derives from direct numerical simulations at low Reynolds number~\citep{Robinson91} and, as a result, our understanding is primarily kinematic. Here the approach is dynamic, in the sense that any form of control implies the selective response of a flow to forcing.

We conclude this section with an acknowledgement that despite the importance of linear mechanisms in wall-bounded turbulence, much research has also focussed on the importance of nonlinear effects. Notable examples include the role that nonlinear mechanisms play in the transition process~\citep{Pringle10,Pringle12,Cherubini10,Cherubini11} wherein the optimal perturbations differ considerably in terms of structure and energy growth, compared to their linear counterparts.

\subsection{Robust control and uncertainty in fluid flows}
With respect to the preceding discussion, linear approximation of the flow dynamics represents just one source of uncertainty between the actual flow~\eqref{navstokes} and state-space models~\eqref{standard} employed for controller design. It is therefore important to identify and model the other sources of uncertainty, as such information can guide the controller design process. An illustrative robust control problem is shown in~Figure~\ref{Fig2}, where~$\mathcal{K}$ denotes the feedback controller,~$\mathcal{U}$ represents model uncertainty and $\mathcal{P}_\mathrm{gen}$ is the nominal (approximate) model of the `generalised' plant, that is, the linearised dynamical model of the fluid flow, the sensors and actuators, as well as the interconnection structure between the plant and controller.

The generalised plant consists of individual partitions that map the control, nonlinear forcing and model uncertainty disturbance input signals,~$\mathsfbi{u}$,~$\boldsymbol{f}$ and~$\mathsfbi{w}$, respectively, to the measured output signal,~$\mathsfbi{y}$, according to:
\begin{subequations}
\begin{equation}\label{Pyf}
	\mathsfbi{y}=\mathcal{P}_{\mathsfbi{w}}\mathsfbi{w}+\mathcal{P}_{f}\boldsymbol{f}+\mathcal{P}\mathsfbi{u}.
\end{equation}
It is worth noting that the individual partitions are~\emph{transfer function} matrices, obtainable from a Laplace transform of a time-domain model. For example,~$\mathcal{P}$ can be obtained from the Laplace transform of~\eqref{standard} as follows:
\begin{equation}
	\mathcal{P}=\mathsfbi{C}(s\mathsfbi{I}-\mathsfbi{A})^{-1}\mathsfbi{B}+\mathsfbi{D},
\end{equation}
where~$s\in\mathbb{C}$ and~$\mathsfbi{I}$ is the identity matrix.
\begin{figure}
 \centerline{\includegraphics[scale=0.09]{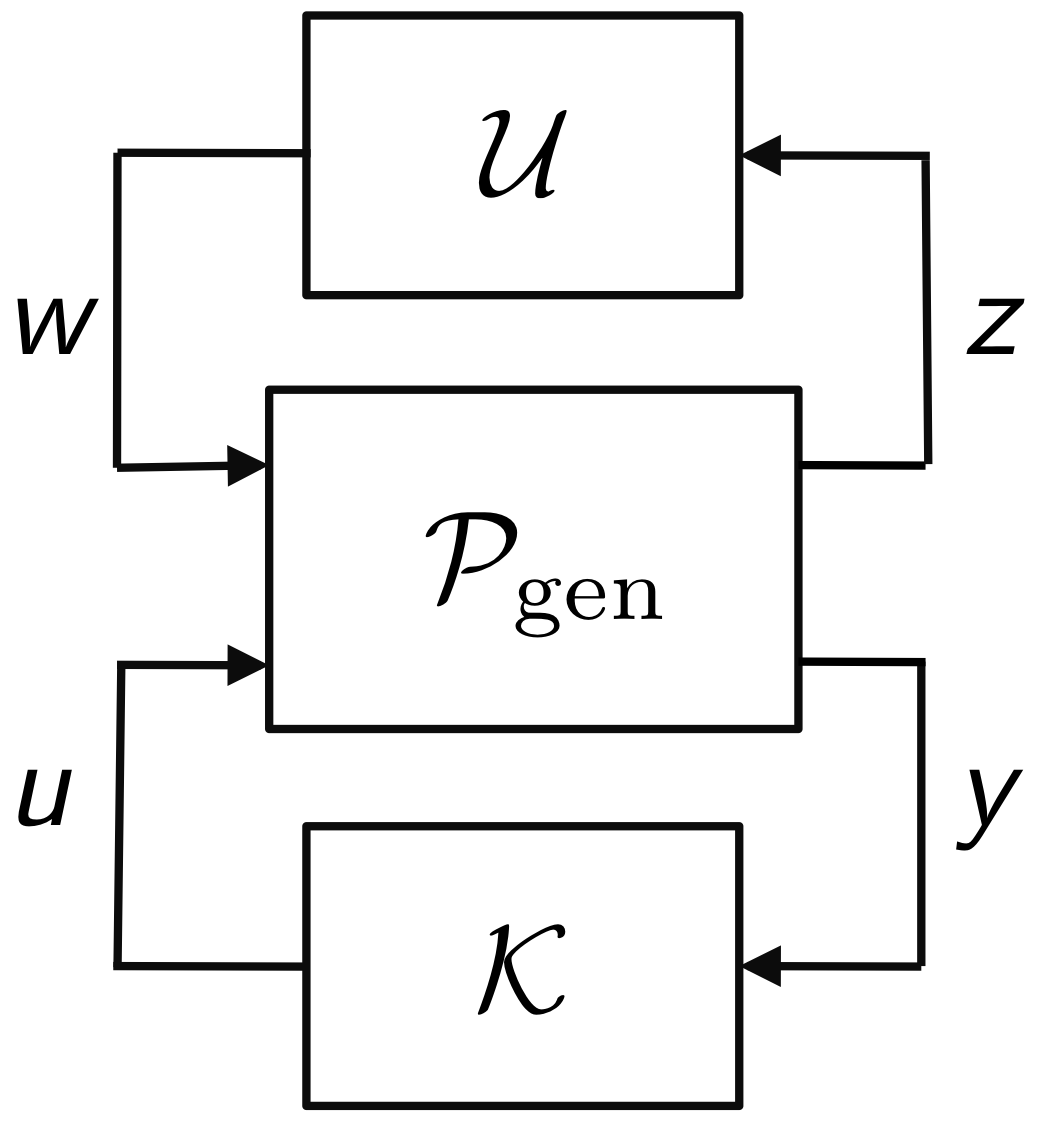}}
  \caption{Robust flow control configuration, including disturbance inputs~$\mathsfbi{w}$ and~$\boldsymbol{f}$ arising from model uncertainty and the nonlinear forcing of the flow, respectively.}
 \label{Fig2}
\end{figure}
The aim of the present work is to design a stabilising controller~$\mathcal{K}$, so as to make the~$\mathcal{H}_\infty$ norm~$\norm{\cdot}_\infty$ of the closed-loop transfer functions from~$\mathsfbi{w}$ and~$\boldsymbol{f}$ to~$\mathsfbi{y}$,~$\mathcal{P}_{\mathsfbi{yw}}$ and~$\mathcal{P}_{\mathsfbi{y}f}$, respectively, both small, where:
\begin{equation}
	\norm{\mathcal{P}_{\mathsfbi{yw}}}_\infty:=\sup_{\mathsfbi{w}\neq0}\frac{\norm{\mathsfbi{y}(t)}_2}{\norm{\mathsfbi{w}(t)}_2},~~~\norm{\mathcal{P}_{\mathsfbi{y}f}}_\infty:=\sup_{\boldsymbol{f}\neq0}\frac{\norm{\mathsfbi{y}(t)}_2}{\norm{\boldsymbol{f}(t)}_2}.
\end{equation}
\end{subequations}
Furthermore, in the interests of robustness, the controller should achieve these aims in the presence of model uncertainty~$\mathcal{U}$, which represents a set of norm bounded transfer function matrices that captures this class of uncertainty. Modelling the uncertainty set again represents a trade-off between complexity and achievable performance, since~$\mathcal{U}$ should be general enough so that the actual plant lies within the set of all perturbed plants defined by the interconnection of~$\mathcal{U}$ with the nominal model~$\mathcal{P}_\mathrm{gen}$, but not so general that closed-loop performance is sacrificed. Within a flow control context, it is desirable that the uncertainty set~$\mathcal{U}$ captures the discrepancy between the actual flow and a simpler control model.~\citet{Bobba} was amongst the first to categorise the uncertainty that arises when approximating~\eqref{navstokes} by a model in linear, state-space form~\eqref{standard}. These sources of uncertainty are summarised as follows:
\newline
\begin{itemize}
	\item~\emph{Model uncertainty.} This takes two forms. The first of these is~\emph{parametric uncertainty} that arises owing to a lack of precise knowledge of the parameters (e.g.~Reynolds number) of the system. Also, if the governing equations are linearised around an equilibrium flow solution, then the theoretical and actual mean flows may differ. Another source of parametric error might arise from numerical errors incurred during the process of eliminating the algebraic constraint~\eqref{navstokesb} to obtain an unconstrained system~\eqref{standarda}. This arises, for example, when inverting an ill-conditioned discretised~Laplacian to obtain the~Orr-Sommerfeld matrix. Secondly,~\emph{dynamic uncertainty}, which is inherent in any finite-dimensional approximation of an infinite-dimensional system. Spatial discretisations of~\eqref{navstokes} only resolve a finite number of dynamic modes, typically those of lowest spatial frequency, and consequently neglect all higher frequency modes. Of those modes that are retained by a spatial discretisation, some will be better resolved than others~\citep{Boyd}. The problem of determining a suitable level of spatial refinement (and hence which modes are of dynamical importance) is of fundamental importance in designing controllers that can tolerate the uncertainty arising from the use of a finite-dimensional flow model. Addressing this issue is an important contribution of this paper.
	\item~\emph{Disturbance uncertainty.} In practice, a flow will be subjected to disturbances arising from a number of sources, such as uncertain boundary conditions, forcing from acoustic noise and the coupling of sensor noise into the flow via a feedback controller. Such disturbances may be impractical to model in any great detail, other than perhaps knowing a bound on their magnitude and the point at which they enter the closed-loop system. In addition, and as discussed in~Section~\ref{Sec1a}, the nonlinearity of the Navier-Stokes equations can be treated as an uncertain disturbance forcing acting upon the linear system. From a control systems perspective this is important, since it enables the problem of suppressing turbulence to be formulated as a disturbance rejection problem. 
\end{itemize}
~\newline
In summary, in order for a feedback controller to guarantee robustness to these sources of model uncertainty, the controller design process must account for each uncertainty in some way. The manner in which this can be achieved is discussed in the following section.

\subsection{Addressing sources of uncertainty}
If bounds on all the uncertainties listed above are known, then each uncertainty can be `extracted' from the plant model to form a~\emph{structured} perturbation matrix~$\mathcal{U}$, and this structure can then be exploited in subsequent controller designs, based on structured-singular-value synthesis algorithms~\citep{Skog}. An alternative, and simpler class of uncertainty model exists in the form of~\emph{unstructured} uncertainty, whereby the perturbation matrix~$\mathcal{U}$ is `full'. Many different unstructured uncertainty models exist~\citep{UncandFeed}, such as~\emph{additive} uncertainty,~\emph{multiplicative input} uncertainty and~\emph{inverse multiplicative output} uncertainty, each with their own merits in terms of representing parametric, dynamic and disturbance uncertainty. An appropriate uncertainty model for closed-loop flow control, for reasons that will be discussed below, is that of~\emph{coprime factor} uncertainty. Background material on this subject is presented in~Appendix~\ref{CFU}, but we note, briefly, that coprime factor perturbations take the form:
\begin{equation}\label{cpfactors}
	\mathcal{P}_{\textrm{p}}:=\left\{(\mathcal{N}+\mathcal{U}_\mathcal{N})(\mathcal{M}+\mathcal{U}_\mathcal{M})^{-1}\right\},
\end{equation}
where the nominal plant~$\mathcal{P}:=\mathcal{NM}^{-1}$ is separated into its stable coprime factors~$\mathcal{N}$ and~$\mathcal{M}$, each of which are perturbed by norm bounded perturbations~$\mathcal{U}_\mathcal{N}$, and~$\mathcal{U}_\mathcal{M}$, respectively, to form a set of perturbed plants~$\mathcal{P}_\mathrm{p}$.

\begin{figure}
  \centerline{\includegraphics[scale=0.3]{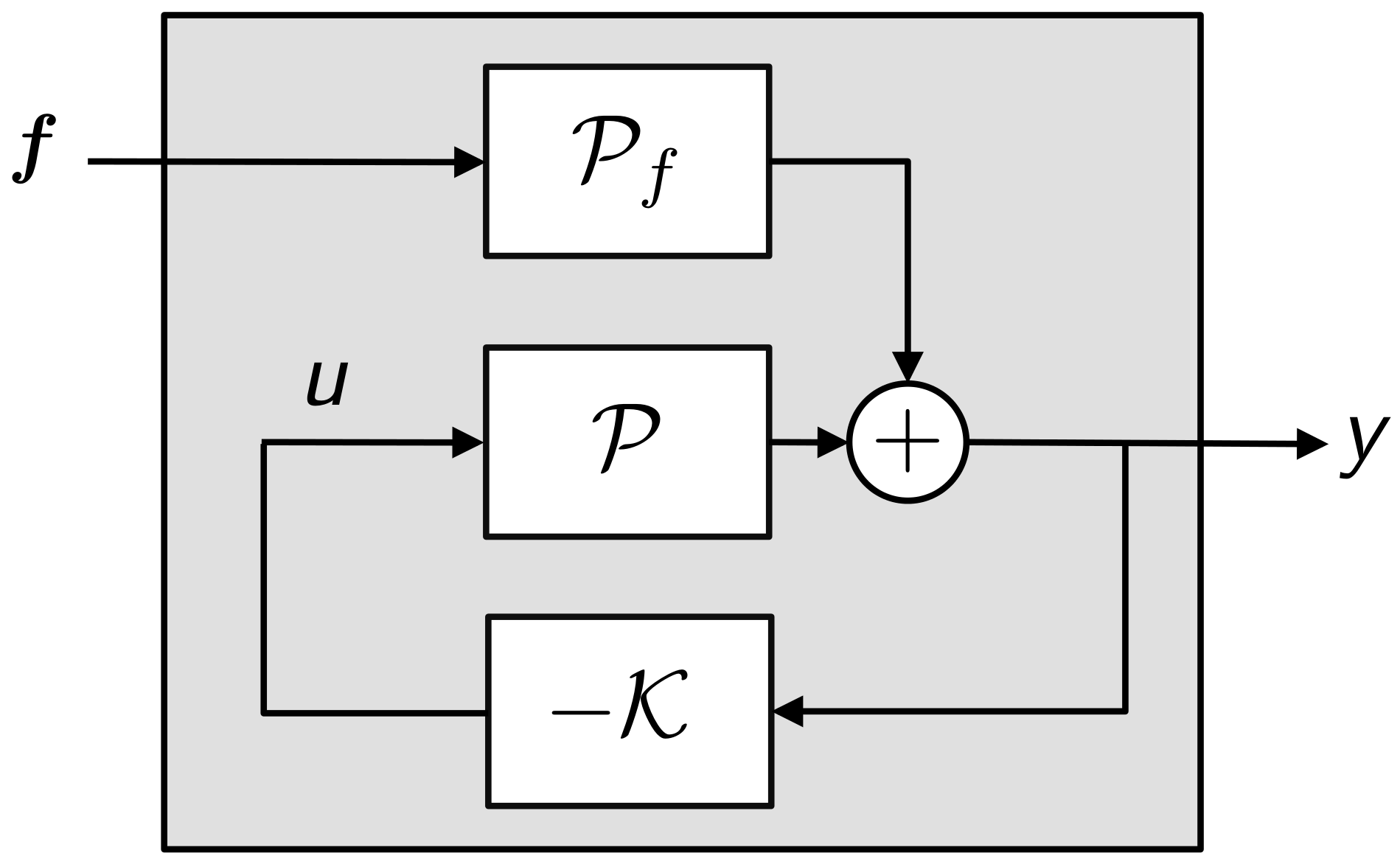}}
  \caption{Feedback control diagram for disturbance rejection. The objective is to design the loop-shaping controller~$\mathcal{K}$ to reject the disturbance, arising form the nonlinear forcing~$\boldsymbol{f}$, upon the measured outputs~$\mathsfbi{y}$ (e.g.~wall shear-stress) of the flow system. The system within the shaded region is the closed-loop transfer function matrix~$\mathcal{P}_{\mathsfbi{y}f}$~\eqref{Pyfcl} from disturbance forcing to output.}
 \label{Fig3}
\end{figure}

Although seemingly abstract, this class of uncertainty is particularly useful as it can be regarded as a blend of multiplicative and inverse multiplicative type uncertainties that naturally account for dynamic and parametric uncertainty, respectively~\citep{UncandFeed}. It also accounts for uncertainty in the number of right-half plane system poles and zeroes, both of which impose fundamental performance limitations upon feedback controllers. It is worth emphasising that the use of such an unstructured uncertainty description greatly reduces the difficulty of modelling the uncertainty set, and hence reduces the difficulty of designing a robust controller. Indeed, in the case of coprime factor uncertainty, no effort is required at all since controller synthesis techniques that employ this description, such as the~$\mathcal{H}_\infty$ loop-shaping procedure of~\citet{McFarlane}, automatically synthesise controllers that maximise the amount of coprime factor uncertainty that a closed-loop system can tolerate. In doing so, and as explained further in~Appendix~\ref{CFU},~$\mathcal{H}_\infty$ loop-shaping controllers also attenuate the effect of disturbances entering at different points in the system (including sensor noise). To see this is the case for rejecting the influence of the forcing arising from the nonlinearity of the flow, consider again the system described by the model~\eqref{Pyf}. Assuming an output feedback control law of the form~$\mathsfbi{u}=-\mathcal{K}\mathsfbi{y}$ leads to the following expression for the closed-loop transfer function~$\mathcal{P}_{\mathsfbi{y}f}$ that relates~$\boldsymbol{f}$ to~$\mathsfbi{y}$:
\begin{equation}\label{Pyfcl}
	\mathsfbi{y}=\underbrace{\left(I+\mathcal{PK}\right)^{-1}\mathcal{P}_f}_{\displaystyle{\mathcal{P}_{\mathsfbi{y}f}}}\boldsymbol{f}.
\end{equation}
The relevant closed-loop system is depicted in~Figure~\ref{Fig3}. The control objective is to reduce the influence of~$\boldsymbol{f}$ upon~$\mathsfbi{y}$, and this is achieved by making the gain of~$\mathcal{P}_{\mathsfbi{y}f}$ small (in terms of~$\|\mathcal{P}_{\mathsfbi{y}f}\|_\infty$), which in turn amounts to designing the loop-shaping controller~$\mathcal{K}$ to ensure that the gain of the open-loop system~$\mathcal{PK}$ is greater than unity, as can be seen from inspection of~\eqref{Pyfcl}.
Such loop-shaping controllers therefore provide a convenient framework for dealing with the parametric, dynamic and disturbance uncertainties encountered when attempting to control flows~\eqref{navstokes} from controllers designed upon simpler models~\eqref{standard}. This simplicity of designing robust controllers has thus meant that~$\mathcal{H}_\infty$ loop-shaping controllers have found use in a variety of applications, ranging from the flight control of vertical take-off aircraft~\citep{Hyde95}, control of combustion oscillations~\citep{Chu03}, bluff body form-drag reduction~\citep{Dahan12} and wind-turbine active blade-pitch control~\citep{Lu14}.

Of the many studies conducted into feedback control of wall-bounded flows, few have explicitly addressed the issue of uncertainty modelling with respect to the class of uncertainties listed above. This is not to say that such feedback controllers have not been robust (at least to some extent), but such assessment has only been possible after closed-loop testing, rather than at the controller design stage. The focus of this paper, therefore, is upon obtaining state-space models~\eqref{standard} of flows described by linearisations of~\eqref{navstokes}, that are of sufficient simplicity to enable straightforward synthesis of controllers with~\emph{a~priori} stability and performance guarantees.

The remainder of this paper is organised as follows. We begin in~Section~\ref{Sec2} by formulating the modelling problem. The starting point is the linearised Navier-Stokes equations and the finishing point is a low-order, state-space model suitable for controller synthesis. On the way we show how to numerically convert a system of DAEs to one of ODEs, and the motivation for doing so. We also introduce the~$\upnu$-gap metric as a useful tool from feedback control theory and show how it can be used to efficiently derive low-order state-space models from spatial discretisations of the linearised flow system. In~Section~\ref{Results}, a~$\mathcal{H}_\infty$ loop-shaping controller is designed from a low-order model and applied to plane channel-flow. Significant portions of this paper are expository in nature and assume little prior knowledge from the reader of feedback control, other than a rudimentary appreciation of classical loop-shaping techniques such as PID control and lead/lag compensation~\citep{Astrom}. To preserve clarity of exposition, some control systems material is included as appendices. In particular, background material on coprime-factor uncertainty and~$\mathcal{H}_\infty$-loop shaping is presented, as are the algorithms employed to firstly convert the semi-discretised Navier-Stokes equations into a standard state-space model.

\section{Formulation of low-order control models}\label{Sec2}
The dynamics of infinitesimal perturbations in a viscous, incompressible, wall-bounded flow can be described by linearisation of the Navier-Stokes equations~\eqref{navstokes} around a mean flow solution. Subsequent spatial discretisation yields a system in the generalised state-space (or descriptor) form:
\begin{align}\label{navstokesdisc}
\underbrace{\begin{bmatrix} \mathsfbi{E}_{11} & 0 \\ 0 & 0 \end{bmatrix}}_{\mathsfbi{E}_\mathrm{D}}
\underbrace{\frac{d}{dt}\begin{bmatrix} \boldsymbol{v}(t) \\ p(t) \end{bmatrix}}_{\dot{\mathsfbi{x}}_\mathrm{D}(t)}=
\underbrace{\begin{bmatrix} \mathsfbi{A}_{11} & \mathsfbi{A}_{12}\\ \mathsfbi{A}_{21} & 0 \end{bmatrix}}_{\mathsfbi{A}_\mathrm{D}}
\underbrace{\begin{bmatrix} \boldsymbol{v}(t) \\ p(t) \end{bmatrix}}_{\mathsfbi{x}_\mathrm{D}(t)}+
\underbrace{\begin{bmatrix} \mathsfbi{B}_1 \\ \mathsfbi{B}_2\end{bmatrix}}_{\mathsfbi{B}_\mathrm{D}}
\mathsfbi{u}(t),
\end{align}
where~$\boldsymbol{v}(t)\in\mathbb{C}^{n_{\boldsymbol{v}}}$ and~$p(t)\in\mathbb{C}^{n_p}$ are the semi-discretised vectors of (perturbation) velocities and pressure, respectively, and~$\mathsfbi{u}(t)\in\mathbb{C}^{m}$ is a vector of control inputs. The state vector is~$\mathsfbi{x}_\mathrm{D}(t)$,~$\mathsfbi{E}_{11}\in\mathbb{C}^{n_{\boldsymbol{v}}\times n_{\boldsymbol{v}}}$ is the symmetric, positive definite mass matrix and~$\mathsfbi{A}_{11}\in\mathbb{C}^{n_{\boldsymbol{v}}\times n_{\boldsymbol{v}}}$ contains a mixture of discrete diffusion and linearised convective terms. The matrices~$\mathsfbi{A}_{12}\in\mathbb{C}^{n_{\boldsymbol{v}}\times n_p}$ and~$\mathsfbi{A}_{21}\in\mathbb{C}^{n_p\times n_{\boldsymbol{v}}}$ represent the discrete gradient and divergence operators, respectively, and~$\mathsfbi{B}_1\in\mathbb{C}^{n_{\boldsymbol{v}}\times m}$ and~$\mathsfbi{B}_2\in\mathbb{C}^{n_p\times m}$ describe how the control inputs influence the states. Note that the subscript~`D', denotes vectors and matrices pertaining to descriptor state-space systems.

The state evolution equation~\eqref{navstokesdisc}, together with the measurement equation~$\mathsfbi{y}(t)=\mathsfbi{C}_\mathrm{D}\mathsfbi{x}_\mathrm{D}(t)+\mathsfbi{D}_\mathrm{D}\mathsfbi{u}(t)$, can be written as a descriptor state-space system:
\begin{subequations}\label{descriptor}
\begin{align}
	\mathsfbi{E}_\mathrm{D}\dot{\mathsfbi{x}}_\mathrm{D}(t)&=\mathsfbi{A}_\mathrm{D}\mathsfbi{x}_\mathrm{D}(t)+\mathsfbi{B}_\mathrm{D}\mathsfbi{u}(t),\label{descriptora}\\
	\mathsfbi{y}(t)&=\mathsfbi{C}_\mathrm{D}\mathsfbi{x}_\mathrm{D}(t)+\mathsfbi{D}_\mathrm{D}\mathsfbi{u}(t),\label{descriptorb}
\end{align}
\end{subequations}
where~$\mathsfbi{E}_\mathrm{D}$,~$\mathsfbi{A}_\mathrm{D}\in\mathbb{C}^{n_\mathrm{D}\times n_\mathrm{D}}$,~$\mathsfbi{C}_\mathrm{D}\in\mathbb{C}^{q\times n_\mathrm{D}}$,~$D_\mathrm{D}\in\mathbb{C}^{q\times m}$ and~$y(t)\in\mathbb{C}^{q}$ is the vector of measured outputs. The order $n_\mathrm{D}=n_{\boldsymbol{v}}+n_p$ of the state vector depends on the resolution, but is typically very large for simulation models (e.g.~$n_\mathrm{D} > 10^6$). For control models, however, the number of states need not be the same, and can in fact be much lower, as discussed in more detail in~Section~\ref{modelref}.

\begin{figure}
  \centerline{\includegraphics[scale=0.08]{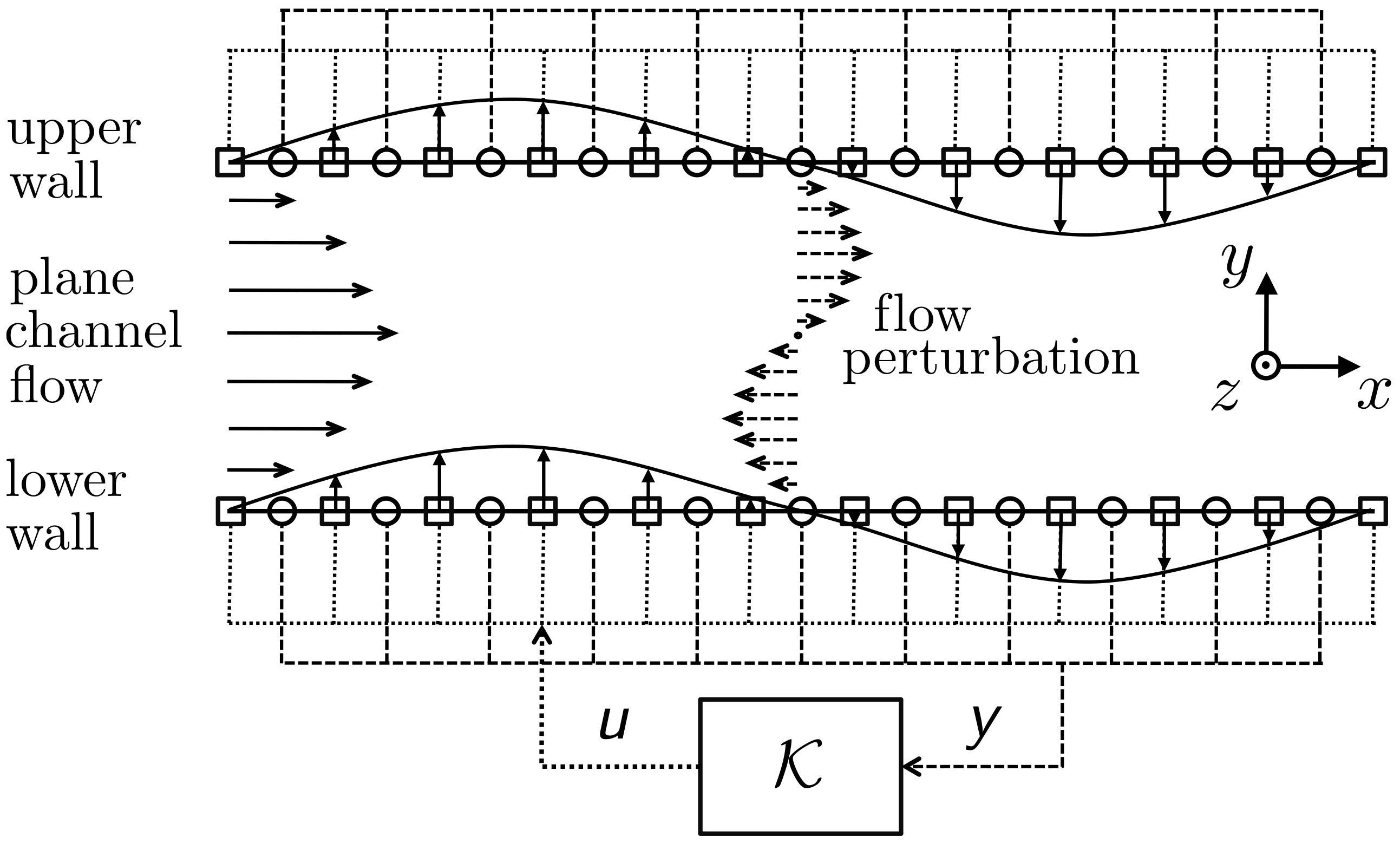}}
  \caption{Side view of plane channel flow and conceptual sketch of the control system. Spatially continuous actuation (transpiration) and sensing (streamwise shear stress) occurs at both walls. For a given wavenumber pair, the feedback controller~$\mathcal{K}$ takes, as inputs, the sensor measurements~$\tilde{\mathsfbi{y}}$, and outputs a control signal~$\tilde{\mathsfbi{u}}$ to the actuators.}
 \label{Fig4}
\end{figure}

As an example of the above formulation, consider the fully developed flow between two infinite, parallel, planar and stationary boundaries, as shown in Figure~\ref{Fig4}.
Non-dimensionalising length scales by the channel half-height, $h$, velocities by the centre-line velocity $U_{\mathrm{cl}}$ and pressure by $\rho{U}^{2}_{\mathrm{cl}}$, the linearised Navier-Stokes equations for incompressible plane channel flow are~\citep{Aamo,McKernanThesis}:
\begin{subequations}\label{channelperturb}
\begin{align}
	\pderiv{u}{t}&=-{U}\pderiv{u}{x}-v\pderiv{U}{y}-\pderiv{p}{x}+\frac{1}{\Rey}\nabla^2 u,\label{channelperturba}\\
	\pderiv{v}{t}&=-{U}\pderiv{v}{x}-\pderiv{p}{y}+\frac{1}{\Rey}\nabla^2 v,\label{channelperturbb}\\
	\pderiv{w}{t}&=-{U}\pderiv{w}{x}-\pderiv{p}{z}+\frac{1}{\Rey}\nabla^2 w,\label{channelperturbc}\\
	0&= \pderiv{u}{x}+\pderiv{v}{y}+\pderiv{w}{z},\label{channelperturbd}
\end{align}
\end{subequations}
where $\Rey=\rho U_\mathrm{cl}h/\mu$ is the Reynolds number and the mean velocity profile satisfies $U=1-y^2$. In non-dimensional co-ordinates the upper and lower walls are located at $y= \pm 1$. The streamwise, wall-normal and spanwise perturbation velocities $u$, $v$ and $w$, respectively, and perturbation pressure $p$ are functions of $x$, $y$, $z$ and $t$ . The initial and boundary conditions are as follows:
\begin{subequations}
\begin{gather}
	u(x,y,z,0)=u_0,\hspace{4mm}
	v(x,y,z,0)=v_0,\hspace{4mm}
	w(x,y,z,0)=w_0,\\
	u(x,\pm 1,z,t)=0,\hspace{4mm}
	v(x,\pm 1,z,t)=0,\hspace{4mm}
	w(x,\pm 1,z,t)=0.\label{noslips}
\end{gather}
\end{subequations}

For the purposes of the current investigation, it is sufficient to employ actuators and sensors that render the system~\eqref{channelperturb} controllable and observable~\citep{Astrom,Bewley98}. Therefore, the walls are assumed continuously distributed with wall transpiration actuators and sensors capable of measuring the streamwise component of the wall-shear stress~\citep{Aamo,Bewley98,McKernan}. A conceptual sketch of this arrangement is shown in~Figure~\ref{Fig4}. The control objective of the present study is to attenuate the streamwise wall-shear stress perturbations. Such a control objective was employed by~\citet{Lee01} and~\citet{Lim03}, where linear controllers were synthesised that significantly reduced the wall-shear stress perturbations, leading to significant reductions in the mean drag.

Actuator dynamics are accounted for by modelling the dynamics of the actuation surfaces as first-order systems with a time constant~$\varsigma\in\mathbb{R}$. For effective control, the actuators should possess sufficient bandwidth to counter the disturbances within the flow. For the present study, a value of~$\varsigma=1$ was found sufficient. Assuming the control inputs~$\mathsfbi{u}(x,y,z,t)$ are voltages supplied to each actuation surface, then the control input can be modelled by the following inhomogenous boundary conditions on the upper and lower walls, respectively:
\begin{subequations}
\begin{gather}\label{ins}
	\pderiv{v(x,+1,z,t)}{t}:=-\frac{1}{\varsigma}v(x,+1,z,t)+\frac{1}{\varsigma}\mathsfbi{u}(x,+1,z,t),\\
	\pderiv{v(x,-1,z,t)}{t}:=-\frac{1}{\varsigma}v(x,-1,z,t)+\frac{1}{\varsigma}\mathsfbi{u}(x,-1,z,t).\label{ins2}
\end{gather}
\end{subequations}
In terms of measurements we consider the streamwise component of the wall shear stress~$\tau_{xy}$ at both walls:
\begin{equation}\label{outs}
	\mathsfbi{y}(x,y,z,t):=\begin{bmatrix}
	\tau_{yx}|_{y=+1}\\
	\tau_{yx}|_{y=-1}
	\end{bmatrix}=\frac{1}{\Rey}\begin{bmatrix}
	\left(\pderiv{u}{y}+\pderiv{v}{x}\right)\!\!\big|_{y=+1}\\
	\left(\pderiv{u}{y}+\pderiv{v}{x}\right)\!\!\big|_{y=-1}
	\end{bmatrix}.
\end{equation}
Having defined inputs and outputs, the infinite-dimensional system~\eqref{channelperturb} is then rendered finite-dimensional via spatial discretisation. The flow is Fourier-transformed in the spatially homogenous $x$ and $z$ directions, in which case the distributed control input~$\mathsfbi{u}$ is approximated as follows:
\begin{equation}
	\mathsfbi{u}(x,\pm 1,z,t)\approx\mathbb{R}\left(\sum_{\mathsf{n}_x=1}^{\mathsf{N}_x}\sum_{\mathsf{n}_z=1}^{\mathsf{N}_z}\tilde{\mathsfbi{u}}(\pm 1,t)e^{i(\alpha x+\beta z)}\right),
\end{equation}
where~$i:=\sqrt{-1}$,~$\alpha$ and~$\beta$ are streamwise and spanwise wavenumbers, respectively, and~$\tilde{\mathsfbi{u}}\in\mathbb{C}^2$ are the Fourier-transformed inputs at each wavenumber pair~$(\alpha,\beta)$. The output equation~\eqref{outs} is similarly approximated. In the inhomogenous $y$ direction the flow is discretised on~$\mathsf{N}_y$ Chebyshev collocation nodes~\citep{Weideman} and the spatial~$y$-derivatives~$\pderiv{}{y},~\pderiv{^2}{y^2}$ are approximated by Chebyshev differentiation matrices~$\mathsfbi{Y}_\mathrm{ch},~\mathsfbi{Y}^2_\mathrm{ch}$, respectively~\citep{Weideman}. Application of the Fourier-transform decouples the system dynamics by wavenumber, and so the flow dynamics for each individual pair~$(\alpha,\beta)$ can now be expressed as a linear, finite-dimensional descriptor state-space system:
\begin{subequations}\label{channeldesc}
\begin{gather}\label{channeldesca}
\underbrace{\begin{bmatrix}
 	\mathsfbi{E}_{\mathrm{D}_{11}} \!\!\!&\!\!\! 0 \!\!\!&\!\! 0 \!\!\!& 0\\
	0 \!\!\!&\!\!\! \mathsfbi{E}_{\mathrm{D}_{22}} \!\!\!&\!\!\! 0 \!\!\!&\!\!\! 0\\
	0 \!\!\!&\!\!\! 0 \!\!\!&\!\!\! \mathsfbi{E}_{\mathrm{D}_{33}} \!\!\!&\!\!\! 0\\
	0 \!\!\!&\!\!\! 0 \!\!\!&\!\!\! 0 \!\!\!&\!\!\! 0
 \end{bmatrix}}_{\displaystyle{\mathsfbi{E}_{\mathrm{D}}}}\underbrace{\frac{d}{dt}\!\!\begin{bmatrix}
 	\tilde{u}_{\mathsf{n}_y}(t)\\
	\tilde{v}_{\mathsf{n}_y}(t)\\
	\tilde{w}_{\mathsf{n}_y}(t)\\
	\tilde{p}_{\mathsf{n}_y}(t)
 \end{bmatrix}}_{\displaystyle{\dot{\mathsfbi{x}}_\mathrm{D}(t)}}\!\!=\!\!\underbrace{\begin{bmatrix}
 	\mathsfbi{A}_{\mathrm{D}_{11}} \!\!\!&\!\!\! \mathsfbi{A}_{\mathrm{D}_{12}} \!\!\!&\!\!\! 0 \!\!\!&\!\!\! \mathsfbi{A}_{\mathrm{D}_{14}}\\
	0 \!\!\!&\!\!\! A_{D_{22}} \!\!\!&\!\!\! 0 \!\!\!&\!\!\! A_{D_{24}}\\
	0 \!\!\!&\!\!\! 0 \!\!\!&\!\!\! \mathsfbi{A}_{\mathrm{D}_{33}} \!\!\!&\!\!\! \mathsfbi{A}_{\mathrm{D}_{34}}\\
	\mathsfbi{A}_{\mathrm{D}_{41}} \!\!\!&\!\!\! \mathsfbi{A}_{\mathrm{D}_{42}} \!\!\!&\!\!\! \mathsfbi{A}_{\mathrm{D}_{43}} \!\!\!&\!\!\! 0
 \end{bmatrix}}_{\displaystyle{\mathsfbi{A}_{\mathrm{D}}}}\underbrace{\begin{bmatrix}
 	\tilde{u}_{\mathsf{n}_y}(t)\\
	\tilde{v}_{\mathsf{n}_y}(t)\\
	\tilde{w}_{\mathsf{n}_y}(t)\\
	\tilde{p}_{\mathsf{n}_y}(t)
 \end{bmatrix}}_{\displaystyle{\mathsfbi{x}_\mathrm{D}(t)}}\!+\!\!\underbrace{\begin{bmatrix}
 	0 \!\!\!&\!\!\!\! 0\\
	\mathsfbi{B}_{\mathrm{D}_{21}} \!\!\!&\!\!\!\! \mathsfbi{B}_{\mathrm{D}_{22}}\\
	0 \!\!\!&\!\!\!\! 0\\
	0 \!\!\!&\!\!\!\! 0
 \end{bmatrix}}_{\displaystyle{\mathsfbi{B}_{\mathrm{D}}}}\!\!\tilde{\mathsfbi{u}}(t),\\
 \tilde{\mathsfbi{y}}(t)=\underbrace{\begin{bmatrix}
 	\mathsfbi{C}_{\mathrm{D}_{11}} & \mathsfbi{C}_{\mathrm{D}_{12}} & 0 & 0\\
	\mathsfbi{C}_{\mathrm{D}_{21}} & \mathsfbi{C}_{\mathrm{D}_{22}} & 0 & 0
 \end{bmatrix}}_{\displaystyle{\mathsfbi{C}_{\mathrm{D}}}}\mathsfbi{x}_{\mathrm{D}}(t),
 \end{gather}
 \end{subequations}
where $\tilde{u}_{\mathsf{n}_y}(t)$,~$\tilde{v}_{\mathsf{n}_y}(t)$,~$\tilde{w}_{\mathsf{n}_y}(t)$ and~$\tilde{p}_{\mathsf{n}_y}(t)$ are vectors containing the Fourier transformed velocity and pressure coefficients at the~$\mathsf{n}_y$-th collocation node (where~$1\leq \mathsf{n}_y \leq \mathsf{N}_y$) for a given wavenumber pair. The elements of the dynamics matrix are defined as:~$\mathsfbi{A}_{\mathrm{D}_{11}}:=\mathsfbi{A}_{\mathrm{D}_{22}}:=\mathsfbi{A}_{\mathrm{D}_{33}}:=-i\alpha U_{\mathsf{n}_y}+\frac{1}{\Rey}\Delta$, $\mathsfbi{A}_{\mathrm{D}_{12}}:=-\frac{d U_{\mathsf{n}_y}}{dy}$, $\mathsfbi{A}_{\mathrm{D}_{14}}:=-\mathsfbi{A}_{\mathrm{D}_{41}}:=-\alpha \mathsfbi{I}$, $\mathsfbi{A}_{\mathrm{D}_{24}}:=-\mathsfbi{A}_{\mathrm{D}_{42}}:=-\mathsfbi{Y}_\mathrm{ch}$, $\mathsfbi{A}_{\mathrm{D}_{34}}:=-\mathsfbi{A}_{\mathrm{D}_{43}}:=-\beta \mathsfbi{I}$,  and $\mathsfbi{E}_{\mathrm{D}_{11}}:=\mathsfbi{E}_{\mathrm{D}_{22}}:=\mathsfbi{E}_{\mathrm{D}_{33}}:=\mathsfbi{I}$, where~$i:=\sqrt{-1}$,~$U_{\mathsf{n}_y}:=1-y_{\mathsf{n}_y}^2$ and~$\Delta:=-\alpha^2+\mathsfbi{Y}_\mathrm{ch}^2-\beta^2$ is the discrete Laplacian operator. The control input influences the states via~$\mathsfbi{B}_{\mathrm{D}_{21}}:=\left[\begin{smallmatrix}\frac{1}{\varsigma} & 0 & \ldots & 0\end{smallmatrix}\right]^T$ and~$\mathsfbi{B}_{\mathrm{D}_{22}}$:=$\left[\begin{smallmatrix}0&\ldots&0&\frac{1}{\varsigma}\end{smallmatrix}\right]^T$. In the case of streamwise shear-stress measurements $\mathsfbi{C}_{\mathrm{D}_{11}}:=\frac{1}{\Rey}\mathsfbi{Y}_{\mathrm{ch}~1,1:\mathsf{N}_y}$ ($1/\Rey$ times the top row of~$\mathsfbi{Y}_\mathrm{ch}$), $\mathsfbi{C}_{\mathrm{D}_{12}}:=\frac{1}{\Rey}\left[\begin{smallmatrix}i\alpha&0&\ldots&0\end{smallmatrix}\right]$,~$\mathsfbi{C}_{\mathrm{D}_{21}}:=\frac{1}{\Rey}\mathsfbi{Y}_{\mathrm{ch}~\mathsf{N}_y,1:\mathsf{N}_y}$, and~$\mathsfbi{C}_{\mathrm{D}_{22}}:=\frac{1}{\Rey}\left[\begin{smallmatrix}0&\ldots&0&i\alpha\end{smallmatrix}\right]$. Boundary conditions~\eqref{noslips}--\eqref{ins2} are enforced in a straightforward fashion by modifying the top and bottom rows of the submatrices in~$\mathsfbi{E}_{\mathrm{D}}$ and~$\mathsfbi{A}_{\mathrm{D}}$. For example, the no slip condition~$u(x,+1,z,t)=0$ is enforced by setting the top rows of~$\mathsfbi{E}_{\mathrm{D}_{11}},~\mathsfbi{A}_{\mathrm{D}_{11}},~\mathsfbi{A}_{\mathrm{D}_{12}}$ and~$\mathsfbi{A}_{\mathrm{D}_{14}}$ equal to zero, except for the~$(1,1)$ element of~$\mathsfbi{A}_{\mathrm{D}_{11}}$ which is set equal to~$1$, whist~\eqref{ins} is satisfied by setting the top rows of~$\mathsfbi{A}_{\mathrm{D}_{22}}$ and~$\mathsfbi{A}_{\mathrm{D}_{24}}$ to zero, with the exception of the~$(1,1)$ element of~$\mathsfbi{A}_{\mathrm{D}_{22}}$, which is set to~$-1/\varsigma$.

\subsection{Dealing with descriptor systems: Eliminating the incompressibility constraint}\label{optimcomp}
Control of descriptor state-space systems~\eqref{descriptor} is less well understood than that for standard state-space systems~\eqref{standard}, and so controller synthesis becomes more straightforward if the former can be converted into the latter. This is trivial when the inverse of~$\mathsfbi{E}_\mathrm{D}$ exists, since both sides of~\eqref{descriptora} can be premultiplied by~$\mathsfbi{E}_\mathrm{D}^{-1}$. However, this is not possible in~\eqref{navstokesdisc} since~$\mathsfbi{E}_\mathrm{D}$ is singular, owing to the assumption of incompressibility. To overcome this difficulty, the system~\eqref{channelperturb} is usually reformulated so that the resulting $\mathsfbi{E}_\mathrm{D}$ matrix is non-singular and can be inverted to yield a standard state-space system. For the case of plane channel flow, it is possible to analytically eliminate the divergence constraint~\eqref{channelperturbd} by reformulating the system in terms of a divergence-free basis described in terms of wall-normal velocities and wall-normal vorticities. A non-singular $\mathsfbi{E}_\mathrm{D}$ can then be obtained by using a set of basis functions that individually satisfy the boundary conditions, yielding the familiar OSS system~\citep{Schmid_Hen,Kim00}:
\begin{subequations}
\begin{align}\label{OSS}
	\frac{d}{dt}
	\begin{bmatrix}\tilde{v}_{\mathsf{n}_y}(t)\\ \tilde{\zeta}_{\mathsf{n}_y}(t)\end{bmatrix}=
	\underbrace{\begin{bmatrix}\mathsfbi{L}_\mathrm{OS} & 0\\ \mathsfbi{L}_\mathrm{C} & \mathsfbi{L}_\mathrm{S}\end{bmatrix}}_{\mathsfbi{A}_\mathrm{OSS}}
	\begin{bmatrix}\tilde{v}_{\mathsf{n}_y}(t)\\ \tilde{\zeta}_{\mathsf{n}_y}(t)\end{bmatrix},
\end{align}
where~$\tilde{\zeta}_{\mathsfbi{n}_y}(t)$ is the vector of~Fourier transformed wall-normal vorticities at a particular wavenumber pair. The~OSS matrix~$\mathsfbi{A}_\mathrm{OSS}$ consists of the Orr-Sommerfeld matrix~$\mathsfbi{L}_\mathrm{OS}$, the Coupling matrix~$\mathsfbi{L}_\mathrm{C}$ and the Squire matrix~$\mathsfbi{L}_\mathrm{S}$:
\begin{align}
	\mathsfbi{L}_\mathrm{OS}&:=\Delta^{-1}\left(-i\alpha U_{\mathsf{n}_y}\Delta+i\alpha\frac{d^2 U_{\mathsf{n}_y}}{dy^2}+\frac{1}{\Rey}\Delta^2\right),\label{OS}\\
	\mathsfbi{L}_\mathrm{C}&:=-i\beta\frac{d U_{\mathsf{n}_y}}{dy},\\
	\mathsfbi{L}_\mathrm{S}&:=-i\alpha U_{\mathsf{n}_y}+\frac{1}{\Rey}\Delta,
\end{align}
\end{subequations}

Although this reformulation has proven itself invaluable for hydrodynamic stability analyses, its use for control system design is not without limitation. For instance, it is difficult to analytically obtain divergence-free bases for more complicated flows, such as those with variable fluid properties, or those with complex geometries~\citep{Ferziger}. This is one of the main reasons why the majority of feedback flows control studies have concentrated on channel flows or similar, parallel, shear flows. Also, satisfying boundary conditions in a divergence-free basis is considerably more difficult than in the original primitive-variable basis, particularly for complex geometries. The boundary conditions, naturally expressed in terms of primitive variables, must be transformed to equivalent conditions in a divergence-free basis that is subject to higher-order spatial derivatives (e.g.~fourth-order in~\eqref{OS}). Failure to satisfy these conditions precisely results in an unphysical system, contaminated by~`spurious' eigenmodes~\citep{Bewley98}.

For these reasons we suggest that the modelling burden is substantially reduced by discretising the flow model and satisfying boundary conditions (in their original, primitive variable form)~\emph{before} converting the resulting finite-dimensional descriptor system~\eqref{descriptor} into standard state-space form~\eqref{standard}. Furthermore, this final step of projecting from descriptor to standard state space form can be performed efficiently via a numerical method~\citep{Schon,Gerdin,Shahzad11}. This is summarised in Appendix~\ref{computesf}, and has been applied successfully to the problem of flow field estimation in a non-parallel boundary layer~\citep{Jones11IJC}. From a high-level perspective, the algorithm takes, as inputs, the matrices of the descriptor system~$(\mathsfbi{E}_\mathrm{D},\mathsfbi{A}_\mathrm{D},\mathsfbi{B}_\mathrm{D},\mathsfbi{C}_\mathrm{D},\mathsfbi{D}_\mathrm{D})$, and outputs the matrices of an equivalent (in the sense that the input-output response is identical) standard state-space system~$(\mathsfbi{A},\mathsfbi{B},\mathsfbi{C},\mathsfbi{D})$, together with a transformation matrix that relates the states~$\mathsfbi{x}_\mathrm{D}$ of the former, to those of the latter~$\mathsfbi{x}$. Applying this algorithm to the system~\eqref{channeldesc} thus yields a standard state-space system of the form~\eqref{standard}:
\begin{subequations}\label{channelsscomp}
\begin{gather}
	\dot{\mathsfbi{x}}(t)=\mathsfbi{A}\mathsfbi{x}(t)+\mathsfbi{B}\tilde{\mathsfbi{u}}(t),\label{channelsscompa}\\
	\tilde{\mathsfbi{y}}(t)=\mathsfbi{C}\mathsfbi{x}(t)+\mathsfbi{D}\tilde{\mathsfbi{u}}(t).
\end{gather}
\end{subequations}

\begin{figure}
  \centerline{\includegraphics[scale=0.5]{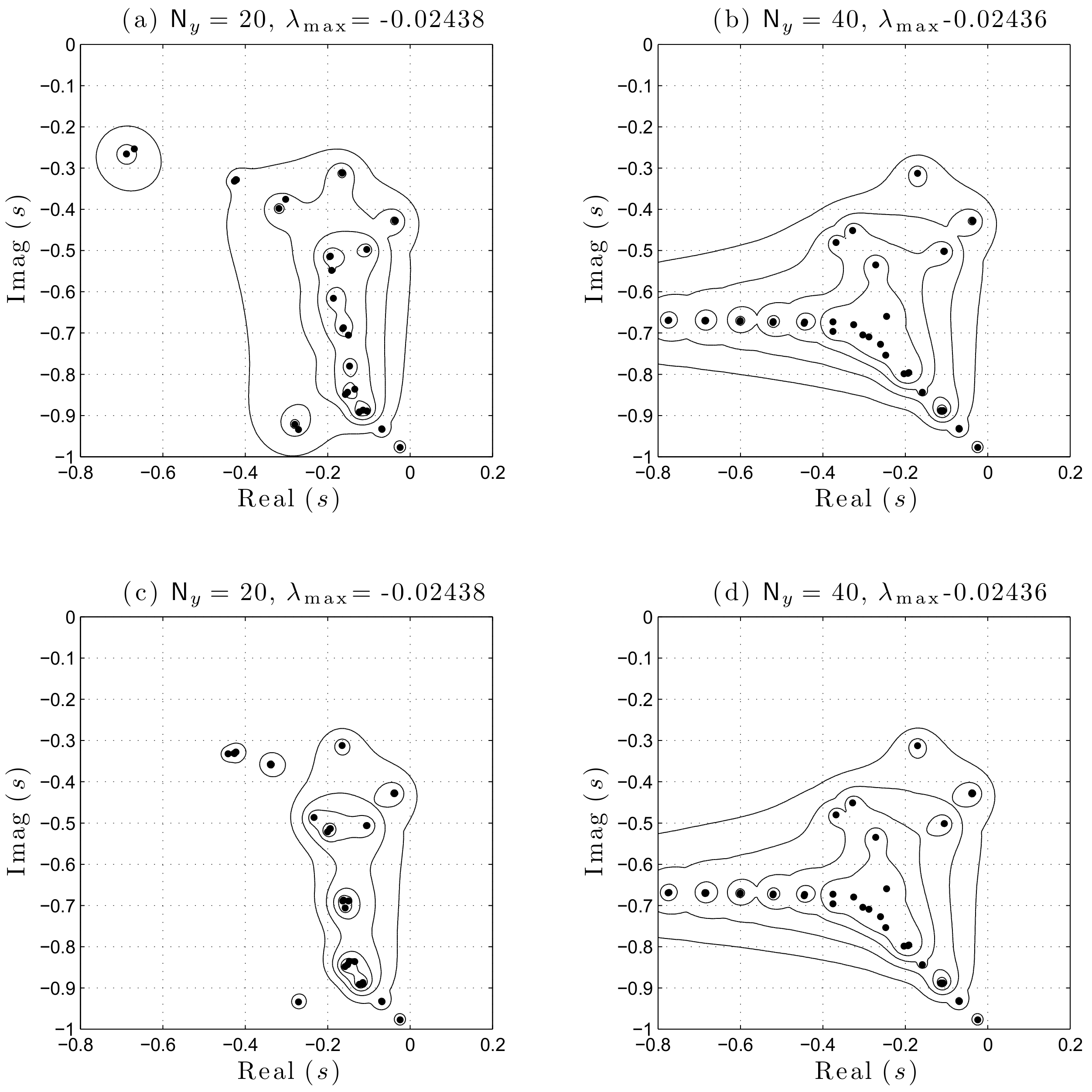}}
  \caption{Eigenvalues (dots) and $\epsilon$-pseudospectra (contours) in lower-left quadrant of the complex plane for (a), (b) $\mathsfbi{A}_\mathrm{OSS}$ in~\eqref{OSS},  and (c), (d) $\mathsfbi{A}$ in~\eqref{channelsscompa}. Pseudospectral contours plotted for~$\epsilon=10^{-3.5}, 10^{-3},\dots, 10^{-2}$ (outermost contour). Left plots are computed for low wall-normal resolution~($\mathsf{N}_{y}=20$ grid-points), right plots are for higher resolution~($\mathsf{N}_{y}=40$). Computed values are for Reynolds number $\Rey=10^3$ and wavenumbers $\alpha=\beta=1$. Also shown are the values of the eigenvalues with maximum real part~$\lambda_\mathrm{max}$, corresponding to those eigenmodes that are least stable in the sense that their eigenvalues are closest to the right-half of the complex plane.}
 \label{Fig5}
\end{figure}

The accuracy of this projection technique can be assessed via a comparison of the spectra and pseudospectra~\citep{TrefethenEmbree} of~$\mathsfbi{A}_\mathrm{OSS}$ in~\eqref{OSS}, with those of the equivalent operator~$\mathsfbi{A}$ in~\eqref{channelsscompa}. Computing the pseudospectra of~$\mathsfbi{A}_\mathrm{OSS}$ in~\eqref{OSS} is complicated by the fact that the kinetic energy of the perturbations is naturally defined in terms of the streamwise, wall-normal and spanwise velocities, thus requiring the energy to be redefined in terms of wall-normal velocity and vorticity (see~\citet{Butler} for details). The eigenvalues and~$\epsilon$-pseudospectra of~$\mathsfbi{A}_\mathrm{OSS}$ and~$\mathsfbi{A}$, for the case~$\Rey=1000,~\alpha=\beta=1$, are shown in~Figure~\ref{Fig5} and begin to show increasing convergence as wall-normal resolution is increased, implying that both operators exhibit the same open-loop transient and asymptotic behaviour. However, an important question to ask is whether or not such reproduction of the open-loop dynamics really matters? Specifically, to what extent does a model employed for~\emph{closed-loop} control need to accurately capture the open-loop dynamics of the actual flow? This issue is discussed in the following section.

\subsection{Modelling for feedback control and the~$\upnu$-gap metric}\label{nugapsec}
As noted by~\citet{Kim07}, a model that is good enough for the purpose of designing a feedback controller, need not necessarily be a good simulation model. However, the converse is also true, in that a good simulation model is not always a suitable model for feedback control design (see e.g.~\citet{Astrom}). It may therefore be misleading to compare the open-loop responses of systems if the objective is to design a feedback controller. This is relevant since most approaches to obtaining low-order models are based on~\emph{open-loop} model-reduction techniques such as balanced truncation~\citep{ZDG96}, proper orthogonal decomposition (POD)~\citep{Holmes96} and balanced POD~\citep{Rowley,Willcox}. Such methods yield models that come with no strict guarantees of being suitable for~\emph{closed-loop} control~\citep{Curtain09}.

\begin{figure}
  \centerline{\includegraphics[scale=0.35]{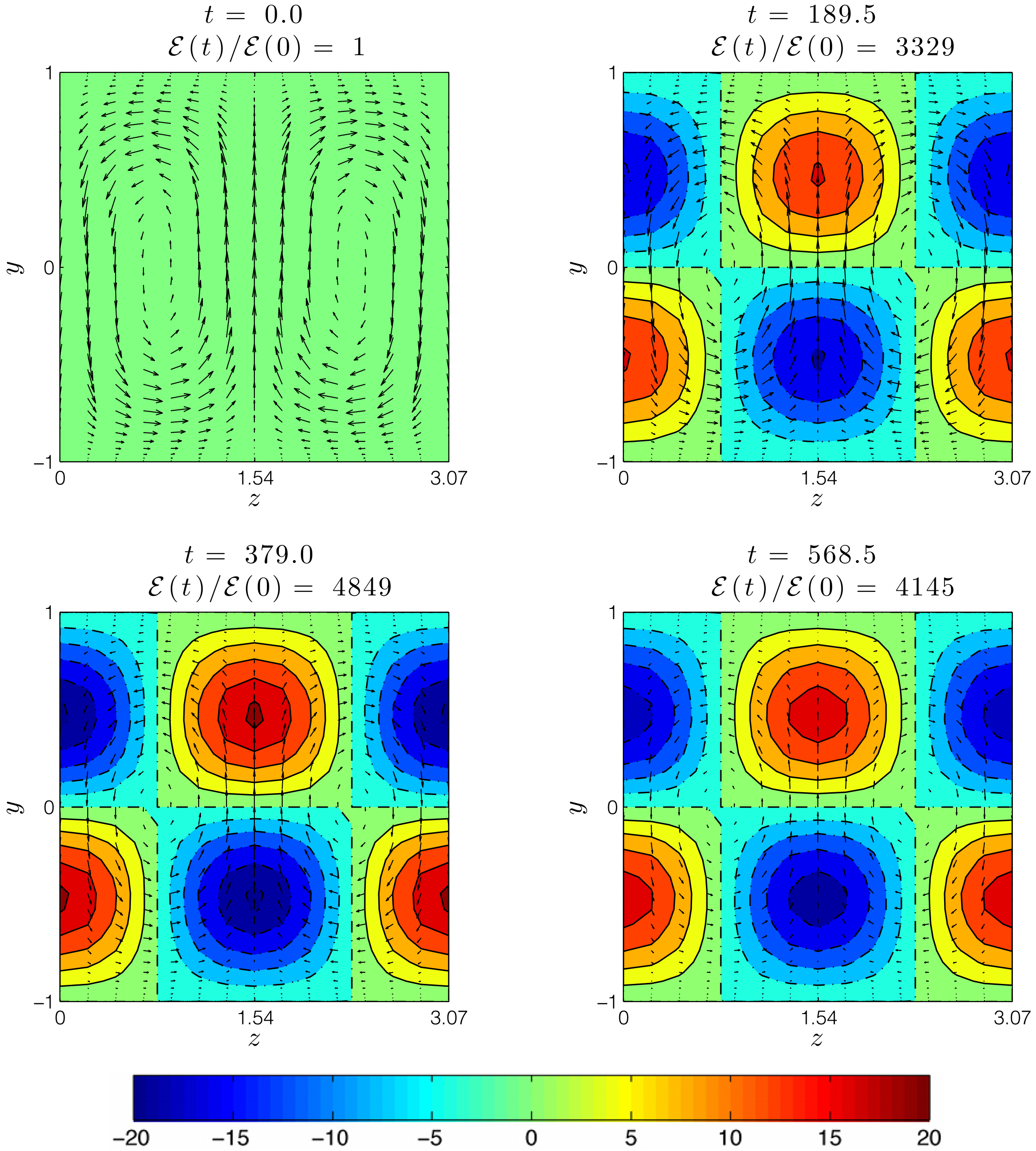}}
  \caption{The loop-shaping design procedure. (a) The nominal flow model~$\mathcal{P}$ is augmented with a precompensator~$\mathcal{W}$ to form a shaped (weighted) plant~$\mathcal{P}_\mathcal{W}:=\mathcal{P}\mathcal{W}$ with a desirable loop-shape. (b) For practical implementation, the precompensator is absorbed back into the controller to form the shaped (weighted) controller~$\mathcal{K}_\mathcal{W}:=\mathcal{WK}$.}
\label{Fig6}
\end{figure}

In order to establish whether or not a model is suitable for feedback control, a measure of `closeness' is required, and fortunately such a measure exists in the form of the~$\upnu$-gap metric~\citep{UncandFeed,Astrom,Zhou}. The definition of the~$\upnu$-gap metric is beyond the scope of the present work, but it suffices to state that the~$\upnu$-gap between two systems, denoted~$\updelta_\upnu(\mathcal{P}_\mathrm{a},\mathcal{P}_\mathrm{b})$, is a metric and thus satisfies the following important properties:
\begin{subequations}
\begin{align}
	&0\leq\delta_\upnu(\mathcal{P}_\mathrm{a},\mathcal{P}_\mathrm{b})\leq 1,\\
	&\delta_\upnu(\mathcal{P}_\mathrm{a},\mathcal{P}_\mathrm{c})\leq\delta_\upnu(\mathcal{P}_\mathrm{a},\mathcal{P}_\mathrm{b})+\delta_\upnu(\mathcal{P}_\mathrm{b},\mathcal{P}_\mathrm{c})~~~\textrm{(Triangle inequality)}.\label{TI}
\end{align}
\end{subequations}
The~$\upnu$-gap assumes systems are connected in feedback by a unity gain controller~($\mathcal{K}=\mathsfbi{I}$). This is a restrictive assumption, but is easily overcome by \emph{shaping} the systems with compensators, as in Figure~\ref{Fig6}, that are designed to shape the open-loop system in a desirable fashion (e.g.~high gain at low frequencies, low gain at high frequencies,~etc.) in a similar manner to classical control methods, such as PID or lag-lead control. The~$\upnu$-gap is then computed between the shaped systems~$\delta_\upnu(\mathcal{P}_{\mathrm{a},\mathcal{W}},\mathcal{P}_{\mathrm{b},\mathcal{W}})$. Thus, the~$\upnu$-gap is very much dependent on the closed-loop objectives encapsulated by the compensator functions. This is important since determining whether or not a model is suitable for designing feedback controllers depends not just on the nominal system, but also upon the closed-loop control objectives. Lastly, the~$\upnu$-gap metric is of considerable practical use in designing~$\mathcal{H}_\infty$ loop-shaping controllers, as explained in more detail in~\citet{UncandFeed}.

Now, suppose~$\mathcal{P}_\infty$ represents the infinite-dimensional flow system obtained from a linearisation of the~Navier-Stokes equations~\eqref{navstokes}, whilst~$\mathcal{P}_\mathsf{n}$ denotes the spatial discretisation of this system upon~$\mathsf{n}$ grid nodes (or finite elements, etc.). The ability to compute~$\updelta_\upnu(\mathcal{P}_\mathsf{n},\mathcal{P}_\infty)$ is important in determining to what extent a spatial discretisation of an infinite-dimensional system yields a suitable model for feedback control. This problem is addressed in the next section.

\subsection{Model refinement and knowing when a spatial discretisation is good enough for closed-loop control}\label{modelref}
One of the main difficulties in designing feedback controllers for fluid flows, based upon finite-dimensional approximations of~\eqref{navstokes}, is deciding what level of spatial discretisation is sufficient. Very fine discretisations are~\emph{likely} to resolve the key dynamics, but the resulting state-space models may be of too great a complexity to enable direct controller synthesis. Model reduction must then be employed to reduce the state-dimension to a more amenable size. Model reduction of large-scale systems~(e.g.~\citet{Antoulas}) is an active research field and various methods exist as mentioned above. Numerical difficulties aside, most of these methods attempt to preserve the open-loop, rather than the closed-loop properties of a system, a choice that may lead to the use of unsuitable models, as discussed in the previous section. Furthermore, and as noted by~\citet{kim03}, most model reduction techniques do not account for the control objective, and yet model `closeness', in a feedback sense, is heavily dependent upon such objectives, as explained previously. It is also important to note that most model reduction techniques attempt to reduce high-dimensional models that in themselves are approximations of an infinite-dimensional system. There is therefore the risk that a control system, designed upon the former, will fail to stabilise the latter, owing to a phenomenon known as `spillover'~\citep{Bal78}, whereby a controller excites unmodelled plant dynamics.

\citet{Jones10} developed an alternative method for obtaining low-order control models of spatially distributed systems, that circumvented each of the problems described above. The method involved computing a sequence of $\upnu$-gaps between low-order plant-models of successively finer spatial resolution,~\emph{starting} from a coarsely discretised (and thus low-order) model. This gradual refinement of model resolution is the conceptual opposite of model reduction-based approaches, and can hence be thought of as~\emph{model refinement}. Generally speaking, as spatial resolution is increased, the sequence of~$\upnu$-gaps between successive plant-models asymptotes towards zero, reflecting the fact that from a closed-loop perspective there are diminishing returns to be obtained from employing highly resolved models. The rate at which the sequence converges to zero is dependent upon the flow, the control objective and the method of spatial discretisation, but can be very great. Establishing the rate of convergence enables the construction of an upper bound on the~$\upnu$-gap between the models in the computed sequence and the infinite dimensional plant, which then informs the selection of a suitable low-order model~\citep{Jones10}. This enables the synthesis, on low-order models, of robust controllers that are guaranteed to stabilise the actual plant, a feature not shared by model reduction methods where the gap between the high-order model (e.g. a DNS flow model) and plant (e.g.~Navier-Stokes equations) is not known, and where the gap between high-order and reduced models may be too expensive to compute. Since the calculation of the bound is based on shaped plant-models of small state-dimension, model refinement avoids the numerical problems inherent in large-scale model reduction-based approaches. Its application here is the first that we are aware of upon a flow control problem.

\begin{figure}
  \centerline{\includegraphics[scale=0.5]{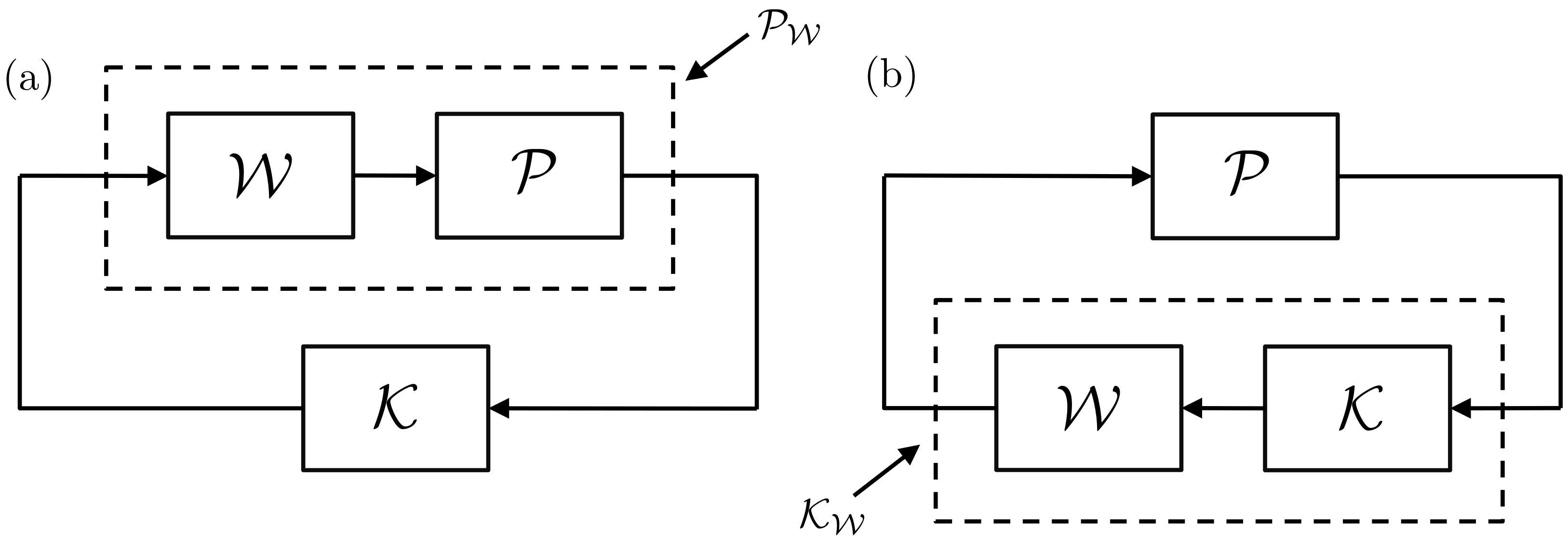}}
  \caption{The model refinement process. The top row shows a sequence of spatially discretised plant-models~$\{\mathcal{P}_{\mathsf{n},\mathcal{W}}\}$, plotted in the space of transfer function matrices, converging to the underlying plant~$\mathcal{P}_{\infty,\mathcal{W}}$ upon refinement of the discretisation. The bottom row shows the construction of the corresponding sequence of~$\upnu$-gaps between the plant-models. The refinement process begins in (a) with the~$\upnu$-gap between a coarsely discretised model~$\mathcal{P}_{\mathsf{n}_i,\mathcal{W}}$ and an incrementally more refined model~$\mathcal{P}_{\mathsf{n}_i+1,\mathcal{W}}$, plotted against the number of grid points~$\mathsf{n}$. In (b) the process is repeated for successively finer discretisations until the~$\upnu$-gaps begin to approach zero. In (c), at some level of refinement~$\mathsf{n}_0$ an analytical sequence is plotted (squares) that upper bounds the~$\upnu$-gap sequence. The summation to infinity of the former provides a bound on the~$\upnu$-gap between the finite-dimensional plant-model~$\mathcal{P}_{\mathsf{n}_0,\mathcal{W}}$ and the infinite-dimensional plant~$\mathcal{P}_{\infty,\mathcal{W}}$ that subsequently informs the design of~$\mathcal{H}_\infty$ loop-shaping controllers.}
\label{Fig7}
\end{figure}

The design procedure is summarised as follows. Firstly, closed-loop objectives are specified by the construction of a precompensator to form the weighted (infinite-dimensional) plant~$\mathcal{P}_{\infty,\mathcal{W}}$. This is then discretised on an initial grid of~$\mathsf{n}_i$ nodes (where~$\mathsf{n}_i$ is small), using an appropriate means of spatial discretisation (finite-difference, finite-element, spectral, etc.), producing a low-order, finite-dimensional plant model~$\mathcal{P}_{\mathsf{n}_i,\mathcal{W}}$. Ideally, one would compute~$\updelta_\upnu({\mathcal{P}_{\mathsf{n}_i,\mathcal{W}},\mathcal{P}_{\infty,\mathcal{W}}})$ directly, but in general this is not possible. However, it is straightforward to form an upper bound as follows. Starting from~$\mathsf{n}=\mathsf{n}_i$, compute the~$\upnu$-gaps between models of successively finer discretisation to form a sequence~$\{\updelta_\upnu({\mathcal{P}_{\mathsf{n},\mathcal{W}},\mathcal{P}_{\mathsf{n}+1,\mathcal{W}}})\}$ and stop when this sequence begins to asymptote towards zero, at some number of grid points~$\mathsf{n}=\mathsf{n}_0$. Then construct a sequence~$\{a_\mathsf{n}\}$ with a finite series (such as a geometric progression) that upper bounds the~$\upnu$-gap sequence for all~$\mathsf{n}\geq \mathsf{n}_0$. The triangle inequality property of the~$\upnu$-gap metric~\eqref{TI} can then be exploited as follows:
\begin{equation}\label{modref}
	\updelta_\upnu({\mathcal{P}_{\mathsf{n}_0,\mathcal{W}},\mathcal{P}_{\infty,\mathcal{W}}})\leq\sum_{\mathsf{n}=\mathsf{n}_0}^{\infty}\updelta_\upnu({\mathcal{P}_{\mathsf{n},\mathcal{W}},\mathcal{P}_{\mathsf{n}+1,\mathcal{W}}})\leq\sum_{\mathsf{n}=\mathsf{n}_0}^{\infty}a_\mathsf{n}.
\end{equation}
Thus, the~$\upnu$-gap between the low-order, finite dimensional plant-model~$\mathcal{P}_{\mathsf{n}_0,\mathcal{W}}$ and the infinite-dimensional plant~$\mathcal{P}_{\infty,\mathcal{W}}$ can be bounded by computing the series of the sequence~$\{a_\mathsf{n}\}$. Then, provided the robust stability margin of a~$\mathcal{H}_\infty$ loop-shaping controller (synthesised from~$\mathcal{P}_{\mathsf{n}_0,\mathcal{W}}$) exceeds this bound by a reasonable margin, then robust closed-loop performance is guaranteed. The assumptions and technical details underpinning this process are fully discussed in~\citet{Jones10}, and a sketch of the procedure is shown in~Figure~\ref{Fig7}.

The model refinement method embodies the fact that sensibly designed feedback control systems are insensitive to unmodelled dynamics occurring at frequencies above the unity gain crossover frequency of the system. These unmodelled dynamics are precisely the type of model uncertainty that arises as a result of spatial discretisation. Referring to~Figure~\ref{Fig5} as an example, the poles and zeros of a spatially distributed system tend to converge to regions in the complex plane upon refinement of spatial discretisation. Finer discretisations resolve dynamics that evolve over shorter timescales, and these correspond to poles further away from the imaginary axis. Sensibly designed loop-shaping controllers show two distinct regions in the frequency domain; low frequencies at which the loop gain is greater than unity, and high frequencies where the gain is less than unity. Generally speaking, the unity gain crossover frequency dictates the region in the complex plane for which the poles and zeros need to be resolved with accuracy. Above this frequency, one can tolerate uncertainty since the loop-gain is less than unity at these frequencies, rendering the closed-loop system insensitive to such uncertainty. Application of the model refinement technique is demonstrated in the following section.

\section{Design of a perturbation wall-shear stress controller}\label{Results}
In this section the model refinement technique is applied to the state-space approximations~\eqref{channelsscomp} of the channel flow perturbation equations~\eqref{channelperturb}, to inform the selection of a low order model from which a~$\mathcal{H}_\infty$ loop-shaping controller is subsequently synthesised. As mentioned in~Section~\ref{Sec2}, the control inputs to the flow system are voltage signals applied to wall transpiration actuators, with measurements of the streamwise component of the wall-shear stress. The control objective is to attenuate the streamwise wall-shear stress perturbations. In this example, the channel-flow models for controller design assume a Reynolds number of~$\Rey=5000$, and a single wavenumber pair of~$\alpha=0$ and~$\beta=2.044$. This corresponds to the optimal conditions for transient energy growth~\citep{Butler}, and so allows for comparisons to be made with other controllers that have been designed for this case (e.g.~\cite{Bewley98}). This is described in more detail in the linear simulation results of~Section~\ref{Linsims}. However, we emphasise that the control objective in the present study is not a function of the initial condition, and so the synthesised controllers should not be viewed purely as transition delay controllers, but rather as controllers that, in this particular instance, attenuate shear stress perturbations, regardless of the source of the perturbation (initial condition, turbulent forcing, etc.). Lastly, we make the point that the same design procedures can be applied to design controllers for flows of arbitrary wavenumber.

\subsection{Controller design}\label{contdes}
The state-space model~\eqref{channelsscomp} is transformed into the following transfer function matrix:
\begin{equation}\label{PNy}
	\mathcal{P}_{\mathsf{N}_y}(s):=\mathsfbi{C}(s\mathsfbi{I}-\mathsfbi{A})^{-1}\mathsfbi{B}+\mathsfbi{D},
\end{equation}
where~$\mathcal{P}_{\mathsf{N}_y}$ is the transfer function matrix obtained from a wall-normal discretisation on~$\mathsf{N}_y$ collocation points. 
\begin{figure}
  \centerline{\includegraphics[scale=0.6]{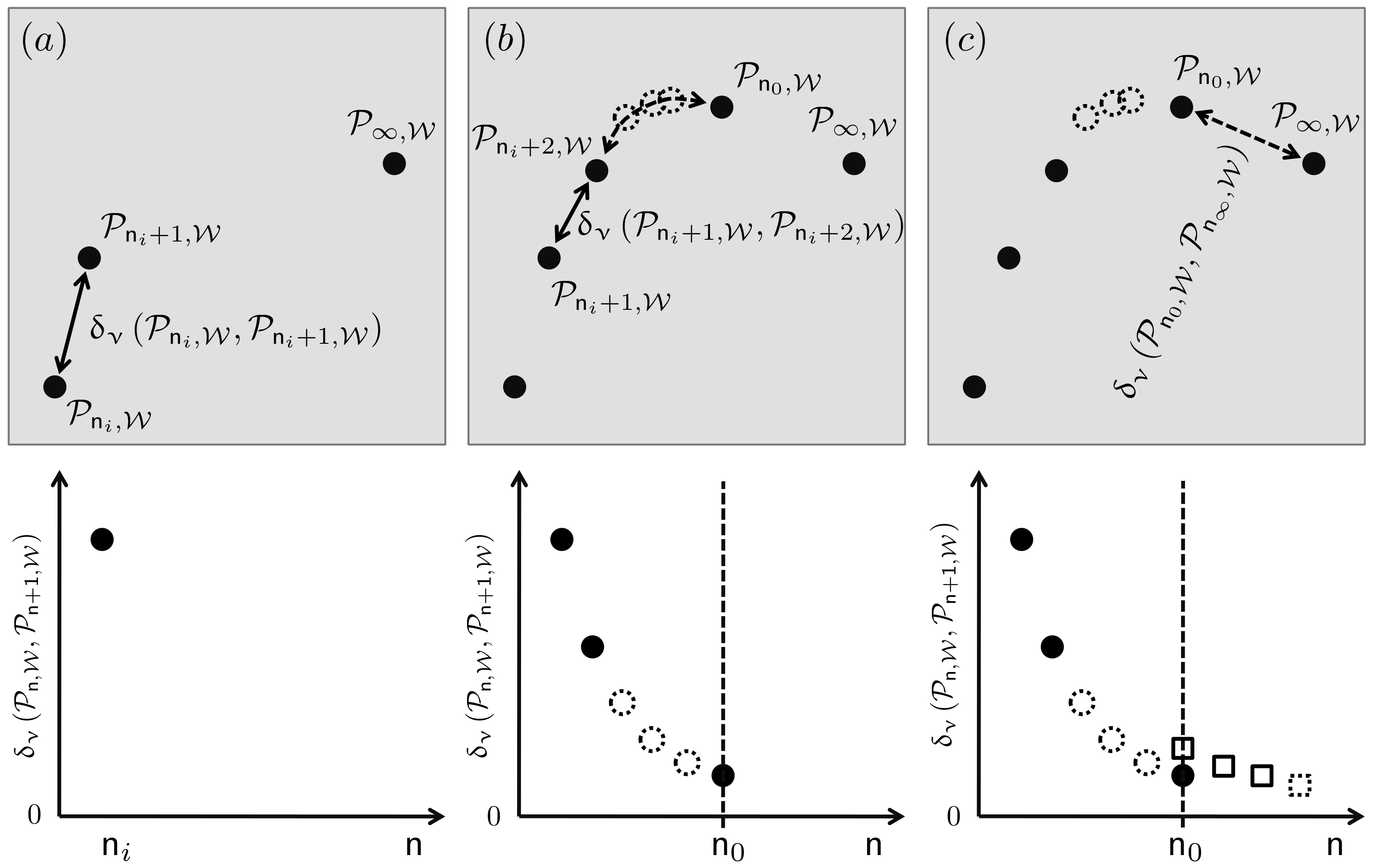}}
  \caption{(a) Open-loop maximum singular value plots of~$\mathcal{P}_{\mathsf{N}_y}$ for~$\mathsf{N}_y=15$~$(\cdot -)$,~$\mathsf{N}_y=50~(-)$, and (b) maximum singular values of the respective compensated system~$\mathcal{P}_{\mathsf{N}_y,\mathcal{W}}$.}
 \label{Fig8}
\end{figure}
The design procedure begins by inspecting the frequency response plots of the maximum singular values of~\eqref{PNy}, denoted~$\bar{\sigma}\left(\mathcal{P}_{\mathsf{N}_y}(i\omega)\right)$. These are plotted in Figure~\ref{Fig8}(a) for two different wall-normal resolutions. Note how the difference in singular value plots for the two different discretisations only becomes pronounced at high frequencies, in this case for temporal frequencies above~$\omega=1$. In terms of control objectives, standard loop-shaping principles are followed in specifying the following design criteria:
\begin{itemize}
	\item Loop crossover frequency at unity gain~$\omega_c\approx 0.3$. Although better performance, in terms of disturbance rejection can be achieved with higher~$\omega_c$, good robustness requires a crossover slope of not much less than~$-1$. Referring to~Figure~\ref{Fig8}(a) the gradient of the singular-value plots decreases rapidly above~$\omega_c\approx 0.3$, as higher frequency poles are encountered, thus limiting the achievable bandwidth of the system.
	\item High loop gain at frequencies below~$\omega_c$. This reduces the effects of disturbances and uncertain parameters in low frequency ranges, noting inparticular that a slope of~$-1$ at~$\omega=0$ provides `integral' control,~i.e.~complete rejection of input disturbances of constant magnitude.
	\item Slope of~$-1$ around~$\omega_c$, for good robustness to coprime factor uncertainty.
	\item Low loop gain at frequencies above~$\omega_c$. This ensures the closed loop is insensitive to noise on sensors, as well as unmodelled high frequency dynamics. Low loop gain is naturally provided by the high frequency poles of the system, but can be augmented with extra poles from the controller, if necessary.
\end{itemize}
These requirements are met by augmenting~$\mathcal{P}_{\mathsf{N}_y}$ with the following precompensator~$\mathcal{W}$:
\begin{equation}
	\mathcal{W}(s):=\begin{bmatrix}
	\frac{2(10s+1)}{s} & 0 & 0 & 0\\
	0 & \frac{2(10s+1)}{s} & 0 & 0\\
	0 & 0 & \frac{2(10s+1)}{s} & 0\\
	0 & 0 & 0 & \frac{2(10s+1)}{s}
	\end{bmatrix},
\end{equation}
Singular value plots of the compensated (weighted) system~$\mathcal{P}_{\mathsf{N}_y,\mathcal{W}}:=\mathcal{P}_{\mathsf{N}_y}\mathcal{W}$ are shown in Figure~\ref{Fig8}(b). Note the greater low-frequency gain, low high-frequency gain, and gentle roll-off at the crossover frequency.

\begin{figure}
  \centerline{\includegraphics[scale=0.65]{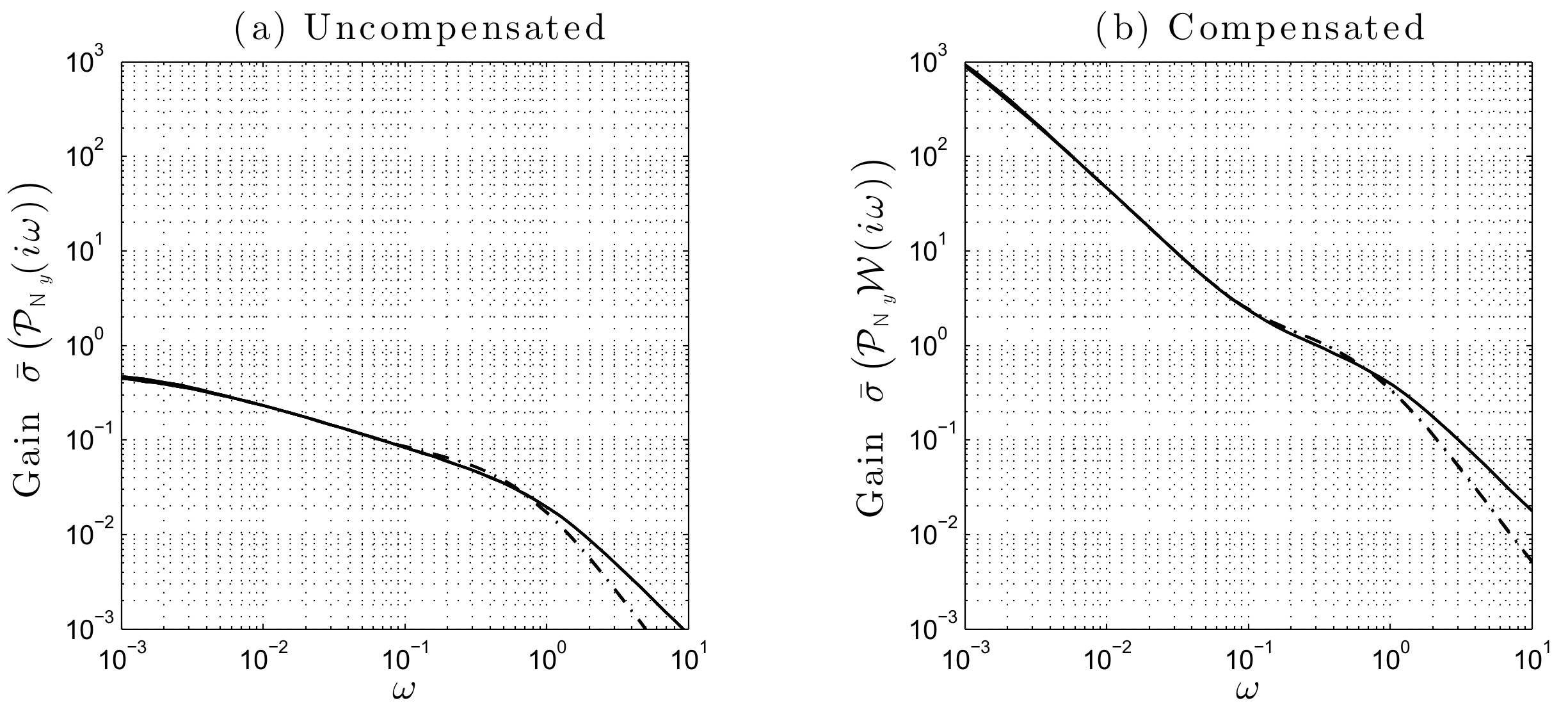}}
  \caption{The model refinement process, showing a plot of (a)~$\updelta_{\upnu}\left(\mathcal{P}_{\mathsf{N}_y,\mathcal{W}},\mathcal{P}_{\mathsf{N}_y+1,\mathcal{W}}\right)$ against grid resolution~$\mathsf{N}_y$, (b)~the same data~$(\cdot)$ plotted on a logarithmic scale, together with a plot~$(\times)$ of the sequence~$\{\mathrm{log}_{10}\left(1.2(0.82)^{\mathsf{N}_y}\right)\}$.}
\label{Fig9}
\end{figure}

Having designed a precompensator, the model refinement procedure is then employed to determine a suitable level of model discretisation. The sequence of~$\upnu$-gaps  between plant models~$\mathcal{P}_{\mathsf{N}_y,\mathcal{W}}$ of successively finer spatial resolution is computed, starting from a low-order model with only~$\mathsf{N}_y=4$ colocation points. The gap between this model, and the next most refined model is~$\updelta_\upnu(\mathcal{P}_{4,\mathcal{W}},\mathcal{P}_{5,\mathcal{W}})=0.69$, which is large and means that a controller designed upon~$\mathcal{P}_{4,\mathcal{W}}$ may not be guaranteed to robustly stabilise~$\mathcal{P}_{5,\mathcal{W}}$, let alone the infinite-dimensional plant~$\mathcal{P}_{\infty,\mathcal{W}}$. However, as the level of discretisation increases, the gaps between models decreases. For example, the gap between~$\mathcal{P}_{30,\mathcal{W}}$ and~$\mathcal{P}_{31,\mathcal{W}}$ is equal to~$0.02$, which is negligible from a robust control perspective. The sequence of~$\upnu$-gaps is plotted in~Figure~\ref{Fig9}(a), from which it is apparent that the~$\upnu$-gaps between successive models rapidly becomes small as model resolution is increased. The same sequence is plotted on a logarithmic scale in~Figure~\ref{Fig9}(b), together with a plot of the following geometric sequence:
\begin{equation}
	\{a_{\mathsf{N}_y}\}:=1.2(0.82)^{\mathsf{N}_y}.
\end{equation}
This sequence forms an upper bound on the~$\upnu$-gap sequence~$\{\updelta_{\upnu}(\mathcal{P}_{\mathsf{N}_y,\mathcal{W}},\mathcal{P}_{\mathsf{N}_y+1,\mathcal{W}})\}$ for~$5\leq \mathsf{N}_y\leq30$. Assuming this holds true for all higher resolutions enables a bound between low-order model and infinite dimensional plant to be computed.
For example, selecting a nominal value of~$\mathsf{N}_y=15$, the following bound on~$\updelta_\upnu(\mathcal{P}_{15,\mathcal{W}},\mathcal{P}_{\infty,\mathcal{W}})$ is obtained from~\eqref{modref}:
\begin{equation}\label{bound}
	\updelta_\upnu(\mathcal{P}_{15,\mathcal{W}},\mathcal{P}_{\infty,\mathcal{W}})\leq\sum_{\mathsf{N}_y=15}^{\infty}1.2(0.82)^{\mathsf{N}_y}=\frac{1.2(0.82)^{15}}{1-0.82}=0.34.
\end{equation} 
It is important to note the bound in~\eqref{bound} was calculated from computations upon low-order models only. This bound was then used to inform the design of a~$\mathcal{H}_\infty$ loop-shaping controller. Direct controller synthesis, based upon the low-order model~$\mathcal{P}_{15,\mathcal{W}}$, yielded a loop-shaping controller~$\mathcal{K}_{15}$ with a robust stability margin of~$b_\mathrm{opt}(\mathcal{P}_{15,\mathcal{W}})=0.68$. A-priori robust performance guarantees (of the controller working well upon the infinite-dimensional plant) are therefore obtained by observing that~$b_\mathrm{opt}(\mathcal{P}_{15,\mathcal{W}})$ exceeds~$\updelta_{\upnu}(\mathcal{P}_{15,\mathcal{W}},\mathcal{P}_{\infty,\mathcal{W}})$ by a reasonable margin of~$0.34$. This is verified by the simulation results in the following sections.

\subsection{Linear simulation results}\label{Linsims}
\begin{figure}
  \centerline{\includegraphics[scale=0.6]{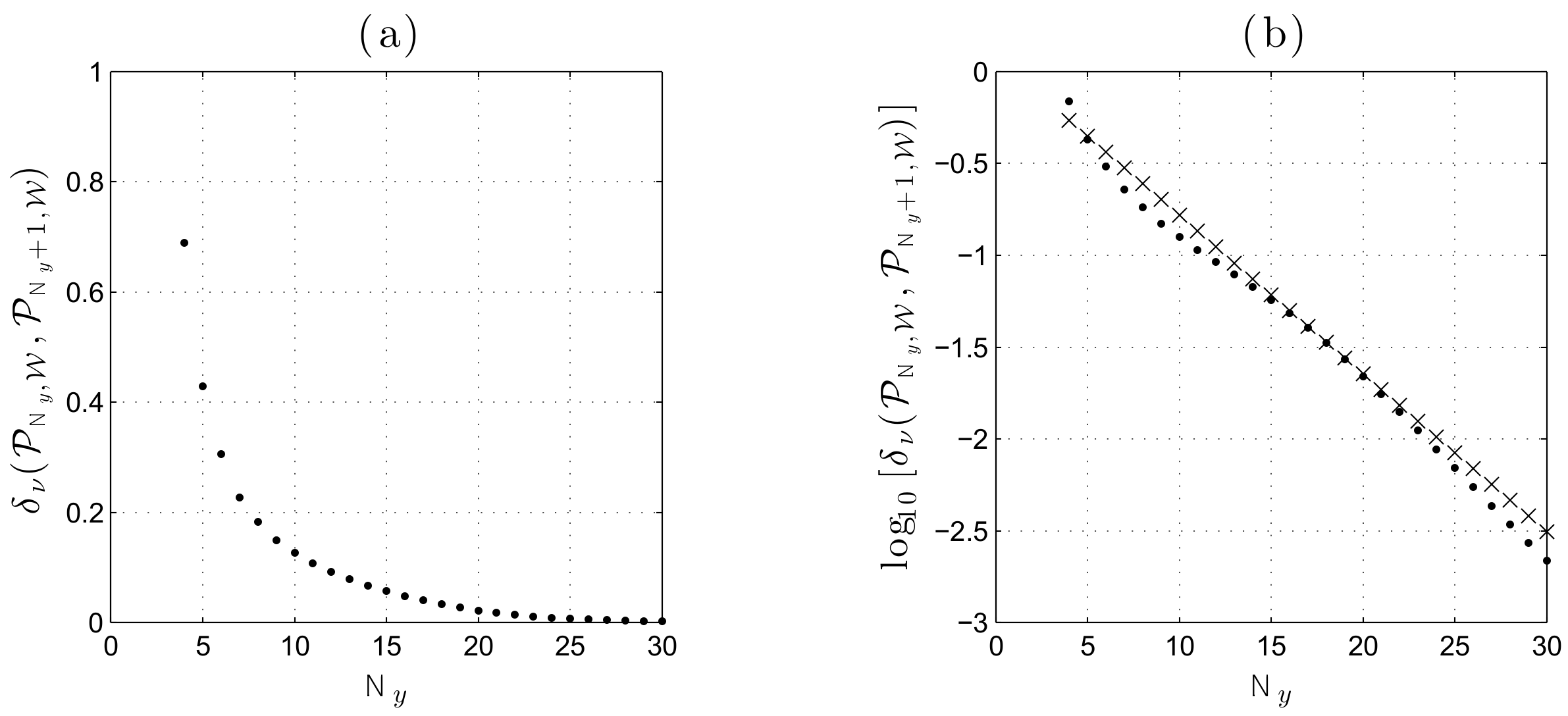}}
  \caption{Linear simulation results. (a) Streamwise wall shear-stresses perturbations against time for the uncontrolled~$(\text{- -})$ and controlled~$(-)$ flows. (b) Perturbation energy gain~$\mathcal{E}(t)/\mathcal{E}_(0)$ against time for the uncontrolled~$(\text{- -})$ and controlled~$(-)$ flows. (c) Controller output~$\tilde{\mathsfbi{u}}(t)$ against time. All control signals are from a weighted controller~$\mathcal{WK}_{15}$ based on~$\Rey=5000$ and~$(\alpha,\beta):=(0,2.044)$.}
 \label{Fig10}
\end{figure}

The controller and precompensator computed in the previous section were combined to form the weighted controller~$\mathcal{WK}_{15}$. This was connected in feedback to a higher-fidelity (linear) flow model~\eqref{channeldesc} employing~$\mathsf{N}_y=100$ wall-normal grid-points, and denoted~$\mathcal{P}_{100}$. The flow was seeded from the optimal initial condition for plane channel flow as computed by~\citet{Butler}, whilst the state vector of the weighted controller was initialised to zero, thus ensuring the controller possessed no prior knowledge of the initial state of the flow.~Figure~\ref{Fig10}(a) shows the evolution of wall-shear stress perturbations~$\tilde{\tau}_{yx}$ against time for both the controlled and uncontrolled flows. After an initial transient period, the perturbations asymptote quickly towards zero under the action of the loop-shaping controller. This is despite the uncertainties arising from the initial state of the controller and from the discretisation error between low and higher-order models employed for controller synthesis and simulation, respectively. In turn, and referring to~Figure~\ref{Fig10}(b), significant attenuation of the perturbation kinetic energy is achieved despite this not being an explicit control objective. The energy gain of the closed-loop system reaches a maximum value of~$\mathcal{E}(t)/\mathcal{E}_0=2857$ at an earlier time of~$t=293$. This represents a~$40\%$ reduction in perturbation energy growth compared to the uncontrolled case. The output from the weighted controller is shown in~Figure~\ref{Fig10}(c). For the sake of comparison, a higher-order controller~$\mathcal{WK}_{35}$ was synthesised and tested on the~$\mathcal{P}_{100}$ flow model. The closed-loop response and control input signal were indistinguishable from those in Figure~\ref{Fig10} obtained from the lower-order controller~$\mathcal{WK}_{15}$. This again underlines the point that spatial refinement of a flow model typically yields diminishing returns in terms of obtaining benefits in closed-loop performance.

\begin{figure}
  \centerline{\includegraphics[scale=0.6]{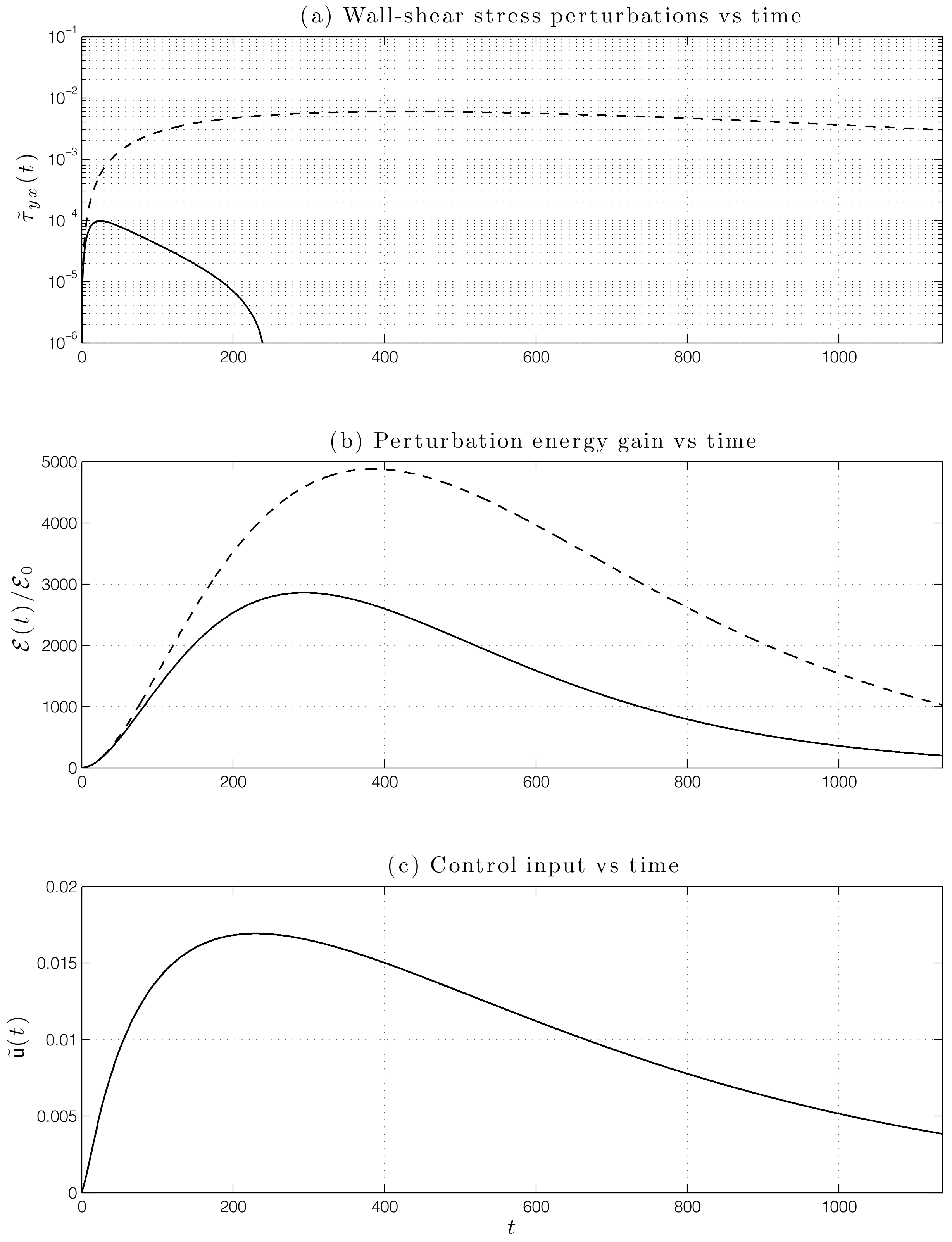}}
  \caption{Velocity components and wall normal derivatives at~$t=293$ for the controlled (--) and uncontrolled~$(\text{- -})$ flows.}
 \label{Fig11}
\end{figure}

\begin{figure}
  \centerline{\includegraphics[scale=0.6]{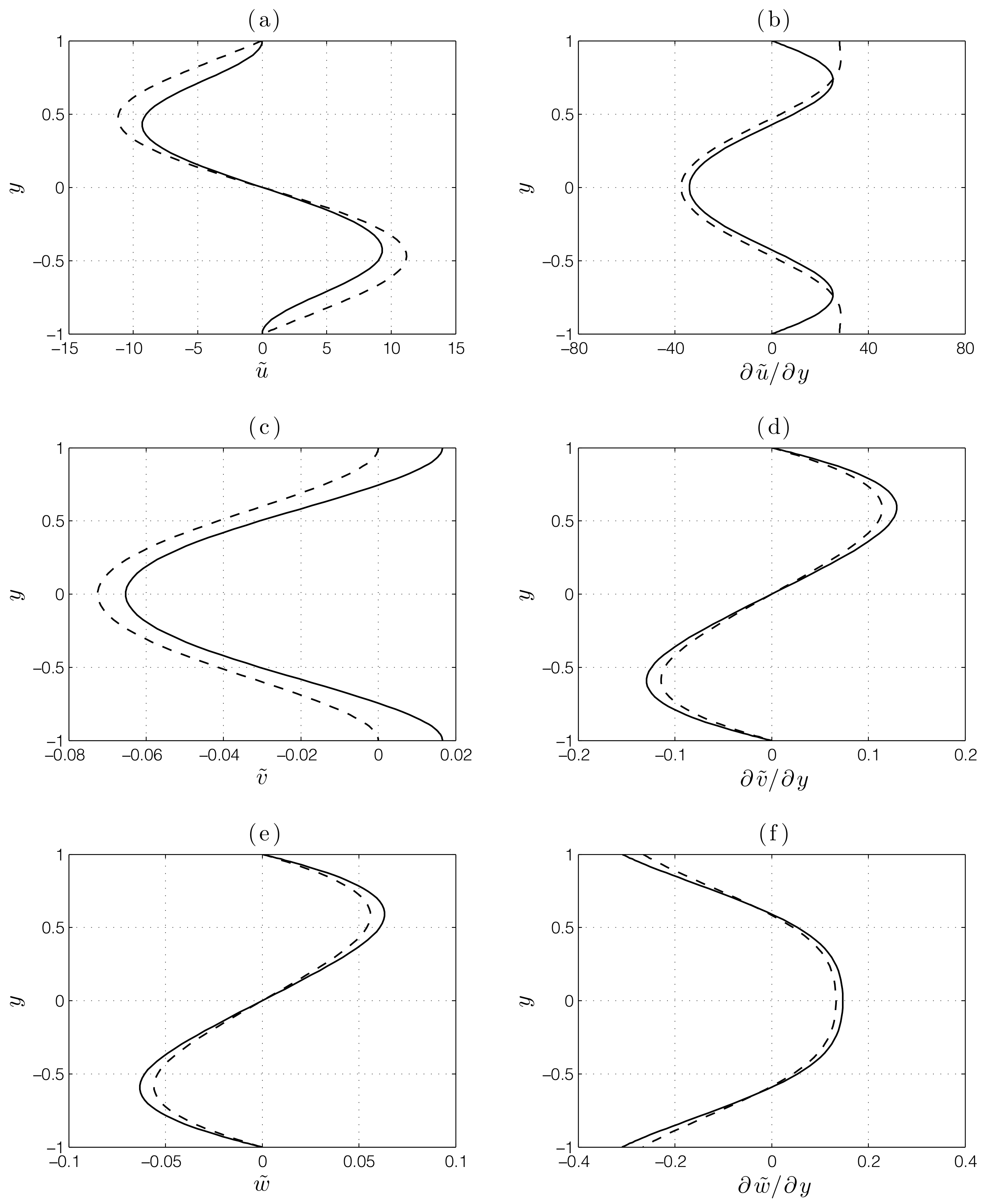}}
  \caption{Evolution of the optimal initial condition in channel flow under the action of the~$\mathcal{H}_\infty$ loop-shaping controller. The controller is designed to attenuate  the magnitude of streamwise perturbation wall shear-stresses. Filled contours represent streamwise perturbation velocities whilst vectors depict the wall-normal and spanwise velocity perturbation fields. Perturbation energy gain~$\mathcal{E}(t)/\mathcal{E}_0$ is also shown.~$\Rey=5000$,~$\alpha=0$,~$\beta=2.044$. Notice the appearance of the buffer vortices close to the walls for~$t>0$.}
\label{Fig12}
\end{figure}

\begin{figure}
  \centerline{\includegraphics[scale=0.55]{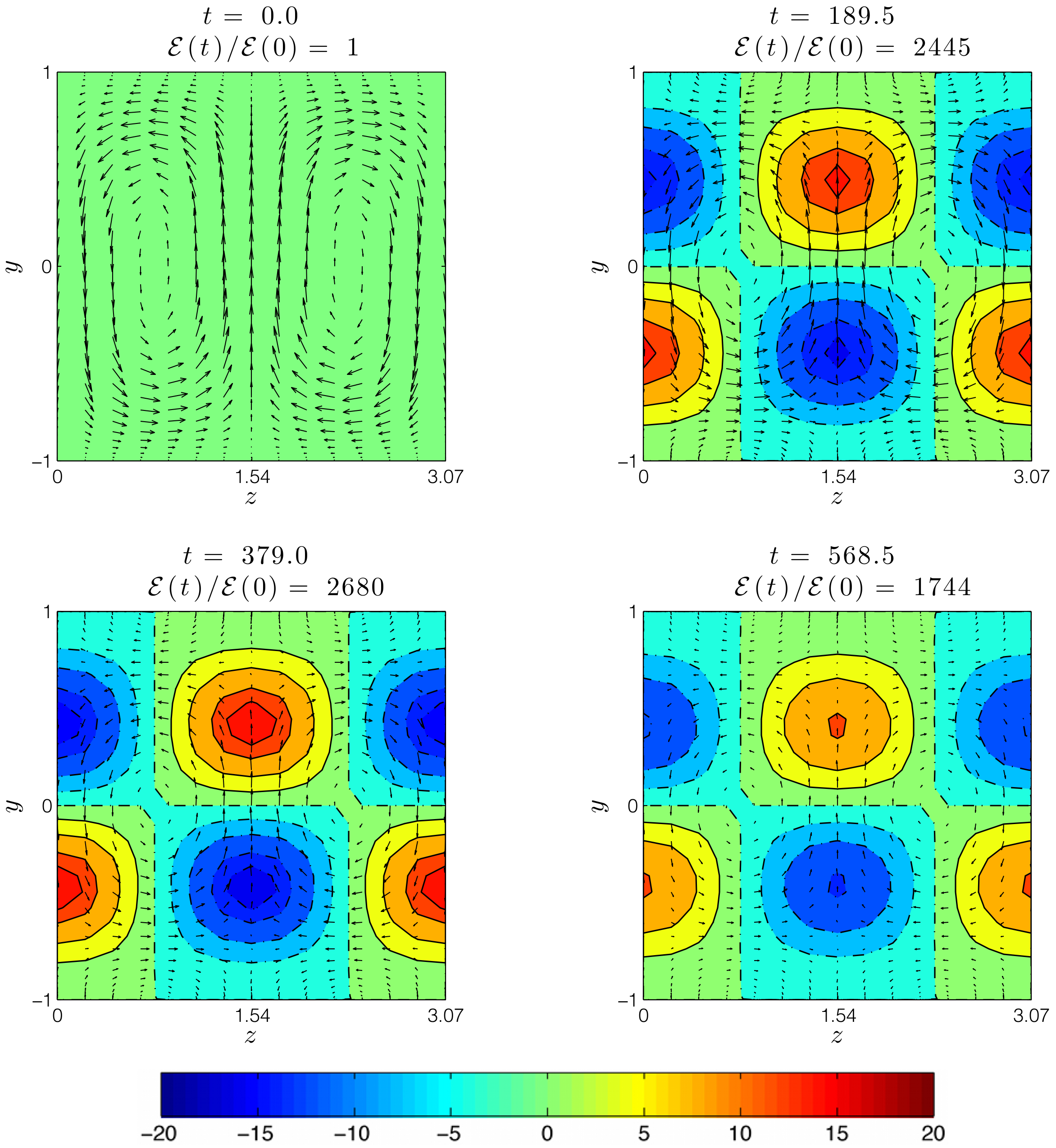}}
  \caption{Streamwise vorticity~$\tilde{\zeta}_x$ at~$t=293$ for controlled (--) and uncontrolled~$(\text{- -})$ flows.}
 \label{Fig13}
\end{figure}

The velocities and their wall-normal derivatives at~$t=293$ are plotted in~Figure~\ref{Fig11}, which illustrates the influence of the controller upon the flow, particularly in the near-wall region. From~Figure~\ref{Fig11}(b) it is clear that the controller has achieved its objective of attenuating the streamwise wall-shear stress perturbations.  Flow visualisations for the controlled case are shown in~Figure~\ref{Fig12}, which demonstrates how the wall transpiration acts to attenuate streak formation, and thus attenuate the perturbation energy. Indeed, the control at the walls creates the small `buffer' vortices observed by~\citet{Bewley98} in their application of transient energy controllers. Such vortices interfere with the shear interaction mechanism that enables velocity perturbations in the channel interior to induce near-wall streaks. A plot of streamwise vorticity against channel height is shown in~Figure~\ref{Fig13} and shows the variation, particularly in the near-wall region between the controlled and uncontrolled flows.

It is interesting to compare the buffer vortices produced by the present controller to those produced by transient energy controllers. Although precise comparisons between different controllers requires the same sensing, actuation and penalty on control effort, the qualitative differences that emerge as a result of employing different control objectives can be inferred. In~\citet{Bewley98}, a~$\mathcal{H}_2$ controller was synthesised upon the same flow as studied here, and achieved a peak energy gain of~$1313$ - approximately half that achieved by the present controller. The buffer vortices induced by the transient energy controller were of sufficient magnitude to produce streamwise streaks that opposed and thus weakened the streaks in the channel centre. However, the presence of these near-wall streaks meant that the streamwise component of the wall shear-stress was considerably greater. Thus the present wall-shear controller can be viewed as a particular case of transient energy controller with a control input penalty of sufficient magnitude to prevent the buffer vortices from inducing opposing near-wall streaks, thus maintaining zero perturbation shear-stress at the wall.

\subsection{Robustness to Reynolds number variations}
For a controller to work well in practice, it must be robust to sources of uncertainty such as model parameter variations. In a flow control context, such variations might include perturbations to the Reynolds number. A controller that offers good performance upon a nominal flow, but destabilises a flow with a slightly different Reynolds number is clearly impractical. Thus, it is important to quantify the performance degradation incurred by attaching a controller to flows with Reynolds numbers different to that employed by the nominal plant model.  Such information can be used to ascertain the range of flows for which a controller will provide acceptable performance. For flows outside this range, other controllers will need to be synthesised, to produce a family of controllers that can be switched between (gain-scheduled) according to the Reynolds number. Again, the~$\upnu$-gap can be used to ascertain bounds on the performance degradation incurred by connecting a nominal controller to a perturbed flow.

\begin{figure}
  \centerline{\includegraphics[scale=0.5]{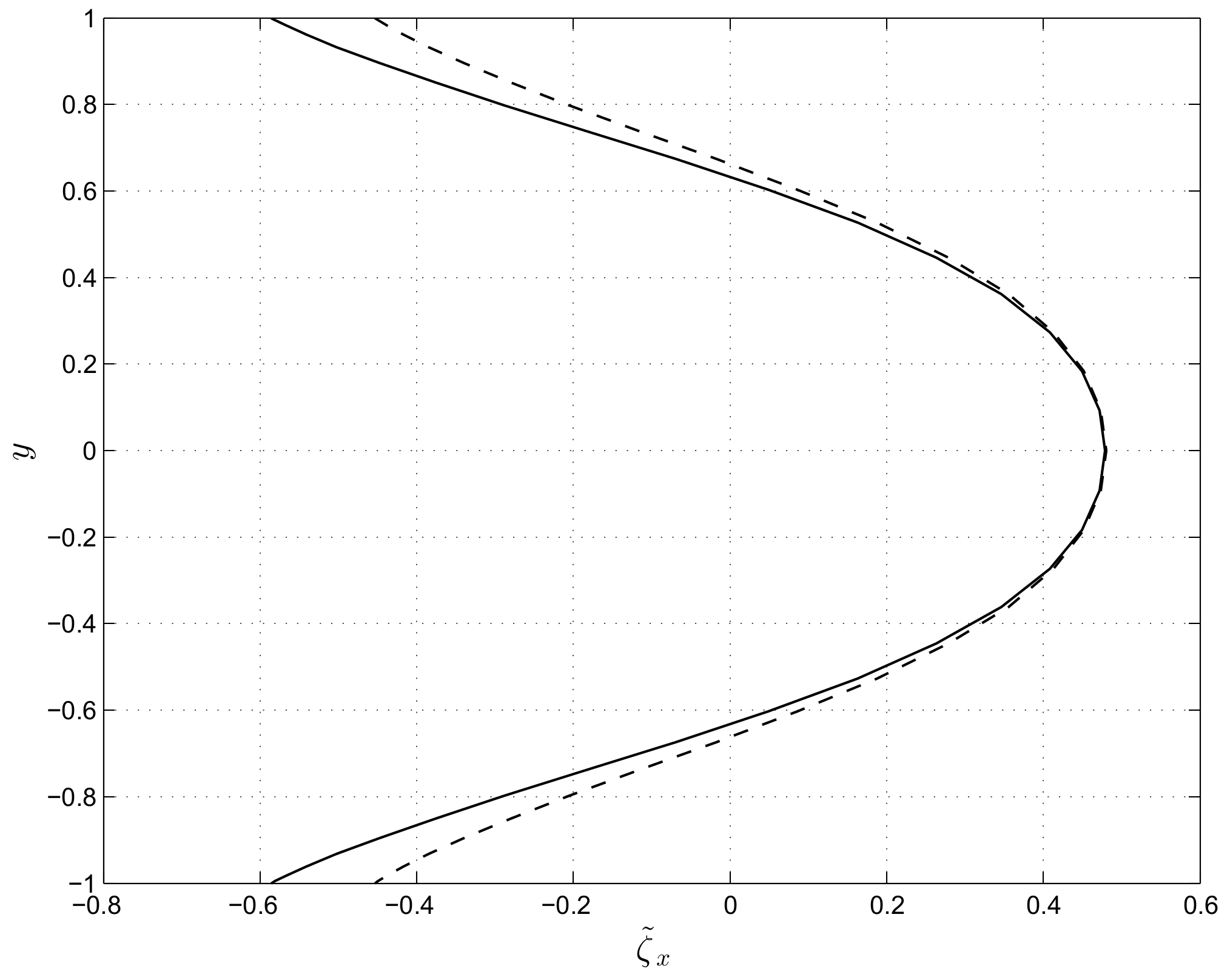}}
  \caption{Variation of the~$\upnu$-gap metric between nominal and perturbed flows. The nominal system is a~$15$ grid-point model based on~$\Rey=5000$,~$\alpha=0$,~$\beta=2.044$. Perturbed models are based on~$100$ grid points and Reynolds numbers in the range~$500\leq\Rey\leq50,000$.}
\label{Fig14}
\end{figure}

\begin{figure}
  \centerline{\includegraphics[scale=0.58]{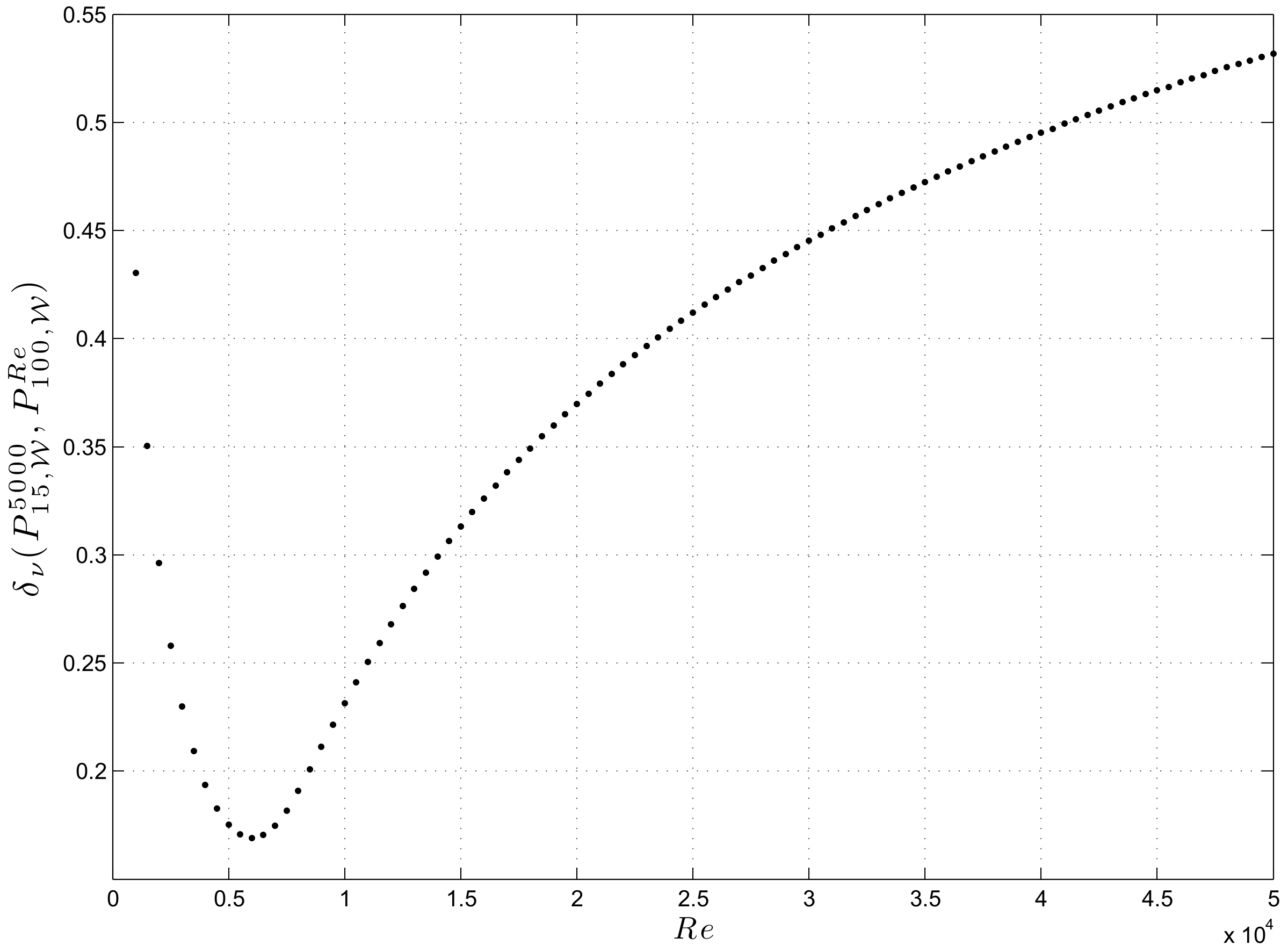}}
 \caption{Streamwise wall-shear stress perturbations against time for the uncontrolled~(- -) and controlled~(--) systems. Here, the controller based on~$\mathcal{P}^{5000}_{15,\mathcal{W}}$ is applied to a perturbed system of higher fidelity and higher Reynolds number,~$\mathcal{P}^{20,000}_{100,\mathcal{W}}$.}
\label{Fig15}
\end{figure}

We begin by computing the~$\upnu$-gaps between the nominal~$15$ grid-point,~$\Rey=5000$ shaped model, denoted~$\mathcal{P}^{5000}_{15,\mathcal{W}}$, and a set of higher fidelity ($100$ grid-point) shaped models at Reynolds numbers in the range~$500\leq\Rey\leq50,000$, denoted~$\left\{\mathcal{P}^{\Rey}_{100,\mathcal{W}}\right\}$. A plot of~$\updelta_\upnu\left(\mathcal{P}^{5000}_{15,\mathcal{W}},\mathcal{P}^{\Rey}_{100,\mathcal{W}}\right)$ is shown in~Figure~\ref{Fig14}.
As expected, the~$\upnu$-gaps are smallest for perturbed flows with Reynolds numbers close to that of the nominal flow and gradually increase as the Reynolds number of the perturbed flows departs from the nominal value. Such information can be used, in conjunction with the controller's stability margin to determine the range of Reynolds numbers over which the nominal controller can be expected to perform well. For example, the robust stability margin of the loop-shaping controller from~Section~\ref{contdes} was computed as~$b_\mathrm{opt}(\mathcal{P}^{5000}_{15,\mathcal{W}})=0.68$. Provided this exceeds the~$\upnu$-gap between the nominal and perturbed flows by a reasonable margin (typically taken to be~$0.3$ - see~Appendix~\ref{CFU} for further details) then one can expect reasonable performance from the controller. The performance requirement is thus~$b_\mathrm{opt}(\mathcal{P}^{5000}_{15,\mathcal{W}})-\updelta_\upnu\left(\mathcal{P}^{5000}_{15,\mathcal{W}},\mathcal{P}^{\Rey}_{100,\mathcal{W}}\right)\geq0.3$ and, referring to~Figure~\ref{Fig14}, this is satisfied for flows in the range~$500\lessapprox\Rey\lessapprox20,000$. One would therefore expect the nominal controller to work well upon~$\Rey=20,000$ flows, and this is confirmed in~Figure~\ref{Fig15}, which shows effective attenuation of the streamwise wall-shear stress perturbations for a linearised flow at this Reynolds number.

\subsection{DNS results}
The results from the previous section were based on a linear model of the flow, and hence neglected the nonlinearity of the~Navier-Stokes equations. In this section, the~$\mathcal{H}_\infty$ loop-shaping controller is tested upon a nonlinear simulation of a channel flow. Nonlinear simulations were performed using a modified version of \emph{Channelflow}, a spectral DNS code for analysis of incompressible Navier-Stokes flow in channel geometries written by \citet{channelflow}. Velocity and pressure are represented as Fourier expansions in the periodic streamwise and spanwise directions and as Chebyshev polynomials in the wall-normal direction. Channelflow uses the influence-matrix method of \cite{Kleiser79} to integrate the Navier-Stokes equations forward in time. This method solves the Navier-Stokes equations at each time step via solutions of a sequence of one-dimensional scalar Helmholtz equations for~$\tilde{u}$,~$\tilde{v}$,~$\tilde{w}$ and~$\tilde{p}$ for each wavenumber pair, with homogenous Dirichlet boundary conditions at the walls. The code was modified in this respect to allow for inhomogeneous boundary conditions to be set at each time step by the controller. The nonlinear terms were computed in skew-symmetric format with 2/3 dealiasing in the streamwise and spanwise directions. A domain size of~$4\pi h\times 2 h \times 1.96\pi h$, in~$x$,~$y$,~$z$ was employed in all testing. The flow field was advanced in time via a third-order semi-implicit backward differentiation algorithm. Testing was performed for a range of Reynolds numbers. Grid resolutions were chosen such that~$\Delta x^+ = 12$,~$\Delta y_\mathrm{min}^{+} = 0.05$ and~$\Delta z^+ = 7$, where $\cdot^+$ notation is used for values expressed in wall units. This led to grid resolutions ranging from~$184 \times 129 \times 158$ grid points in~$x$,~$y$,~$z$ for the~$\Rey_\tau = 175$ case, to~$394 \times 193 \times 338$ grid points in~$x$,~$y$,~$z$ for the~$Re_\tau=360$ case. Initial conditions consisting of small random perturbations to a laminar mean profile were employed to transition the flow to a fully turbulent state. This latter condition was validated by comparing the mean velocity profiles and perturbation root-mean-square profiles from each test case to the benchmark data of~\cite{Moser99} and~\cite{Iwamoto02}. We emphasise that it was only once the flow was fully turbulent (at a nominal time~$t=0$) that the controllers were activated.

\begin{figure}
  \centerline{\includegraphics[scale=0.5]{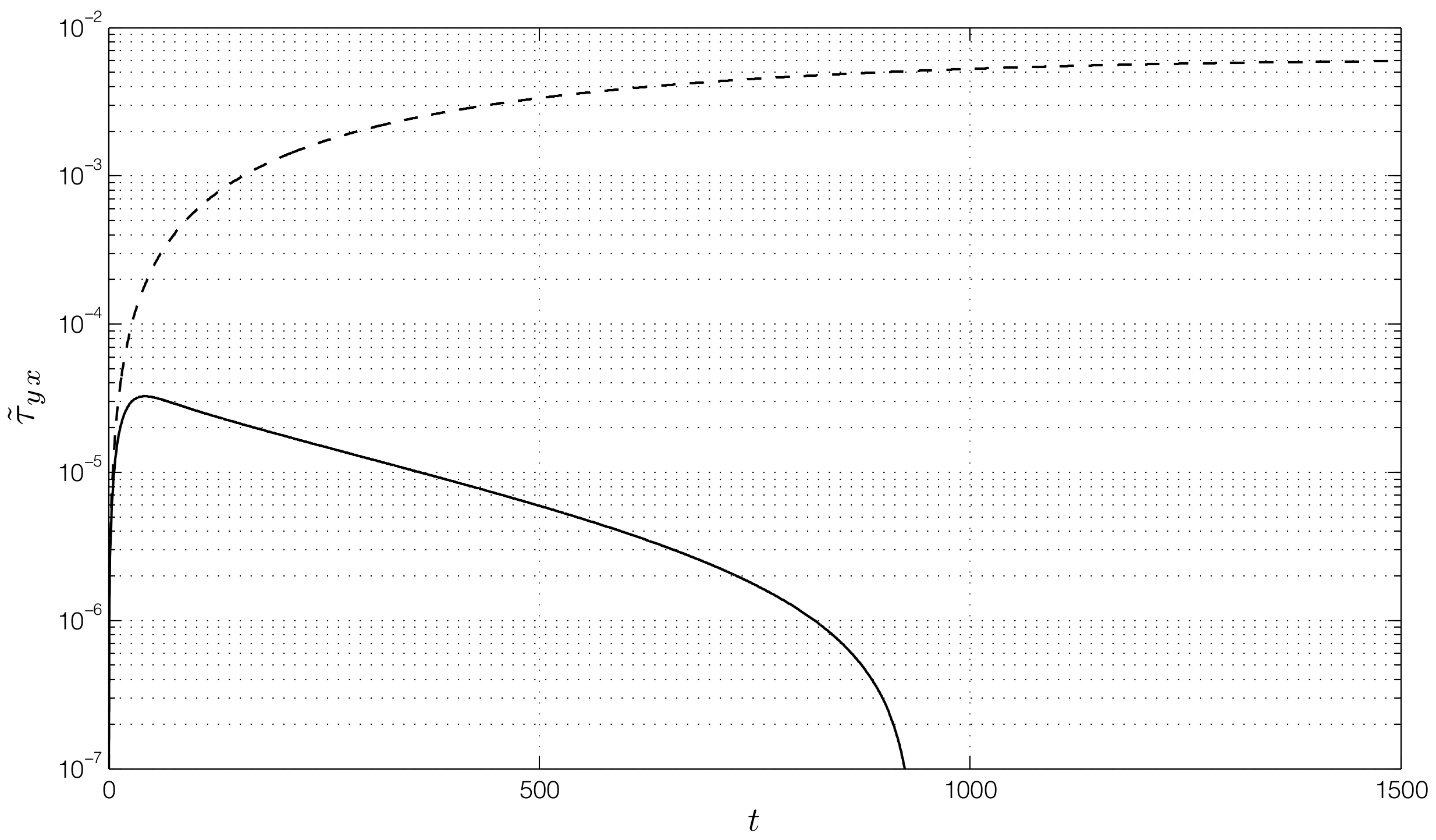}}
  \caption{Time evolution of the magnitude of the wall-shear stress perturbations for uncontrolled~(- -) and controlled~(--) cases. Values are normalised by the wall-shear stress magnitude at the time when the controller is switched on~$(t=0)$.}
\label{Fig16}
\end{figure}

\begin{figure}
  \centerline{\includegraphics[scale=1.0]{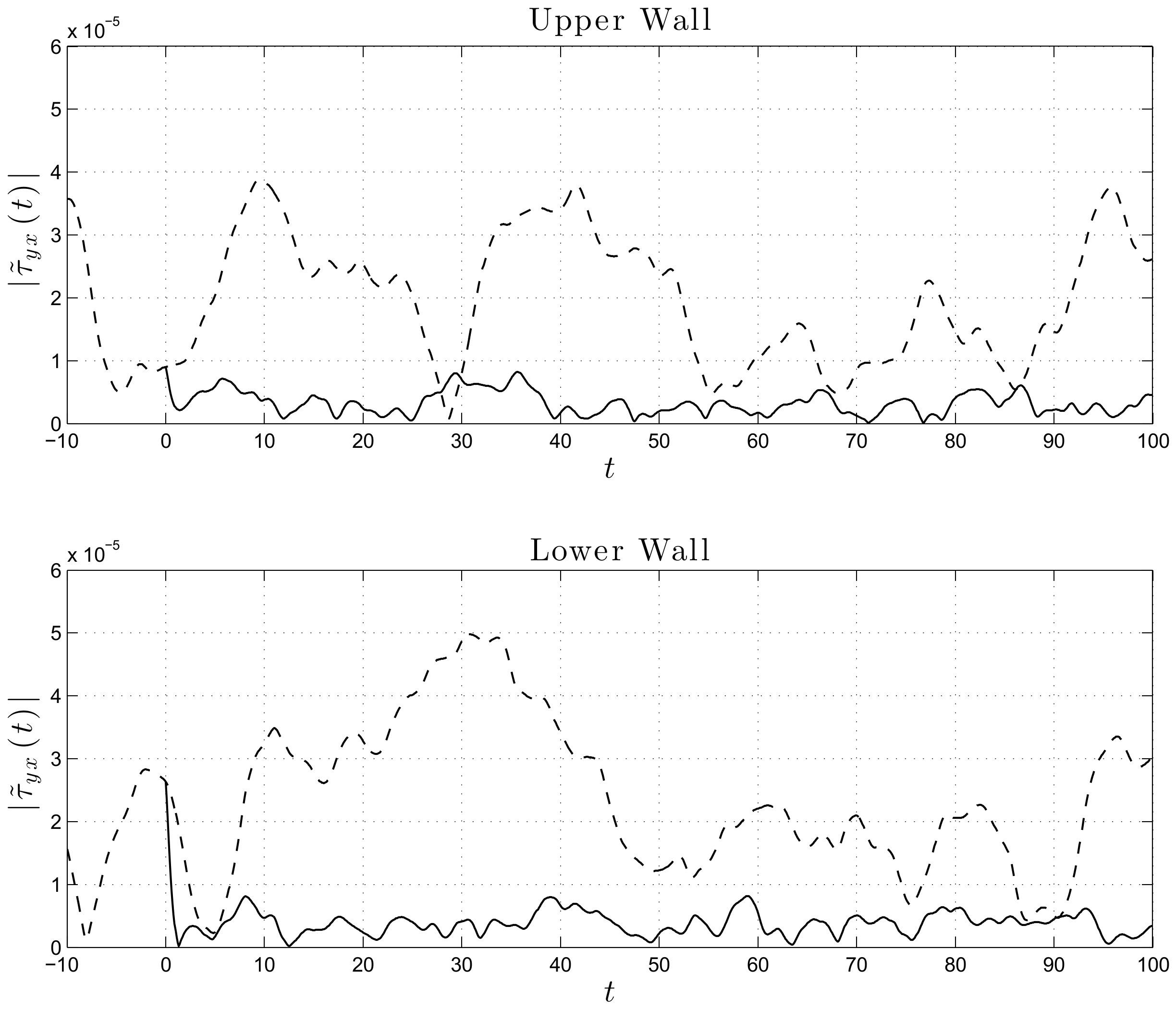}}
  \caption{Representative snapshot of the controlled (top) and uncontrolled (bottom) flows at the lower wall,~taken at~$t=1000$. The figure shows contours of perturbation streamwise velocity (shaded regions) and perturbation streamwise vorticity (solid and dashed lines) in wall units. The near-wall streaks are significantly attenuated by the controller. The near-wall buffer vortices induce just enough streak formation of reverse sign to reduce the wall shear stress.}
\label{Fig17}
\end{figure}

The~$\mathcal{H}_\infty$ loop-shaping controller was applied to a~$\Rey_\tau=210$ flow. Figure~\ref{Fig16} shows the effect of this controller upon the magnitudes of the streamwise perturbation wall-shear stresses, computed for wavenumber pair~$(\alpha,\beta)=(0,2.044)$. The controller was activated at time~$t=0$ and quickly acted to attenuate the wall shear-stress perturbations at both walls, achieving an~$87\%$ reduction in the RMS wall-shear stress perturbations. Snapshots of the controlled and uncontrolled flows are displayed in~Figure~\ref{Fig17}, from which it is evident that the near-wall streaks are significantly attenuated by the action of the controller. The figure shows the appearance of buffer vortices extremely close to the wall, created by the controller. They are of just sufficient amplitude to attenuate the wall shear stress.

\begin{figure}
  \centerline{\includegraphics[scale=0.5]{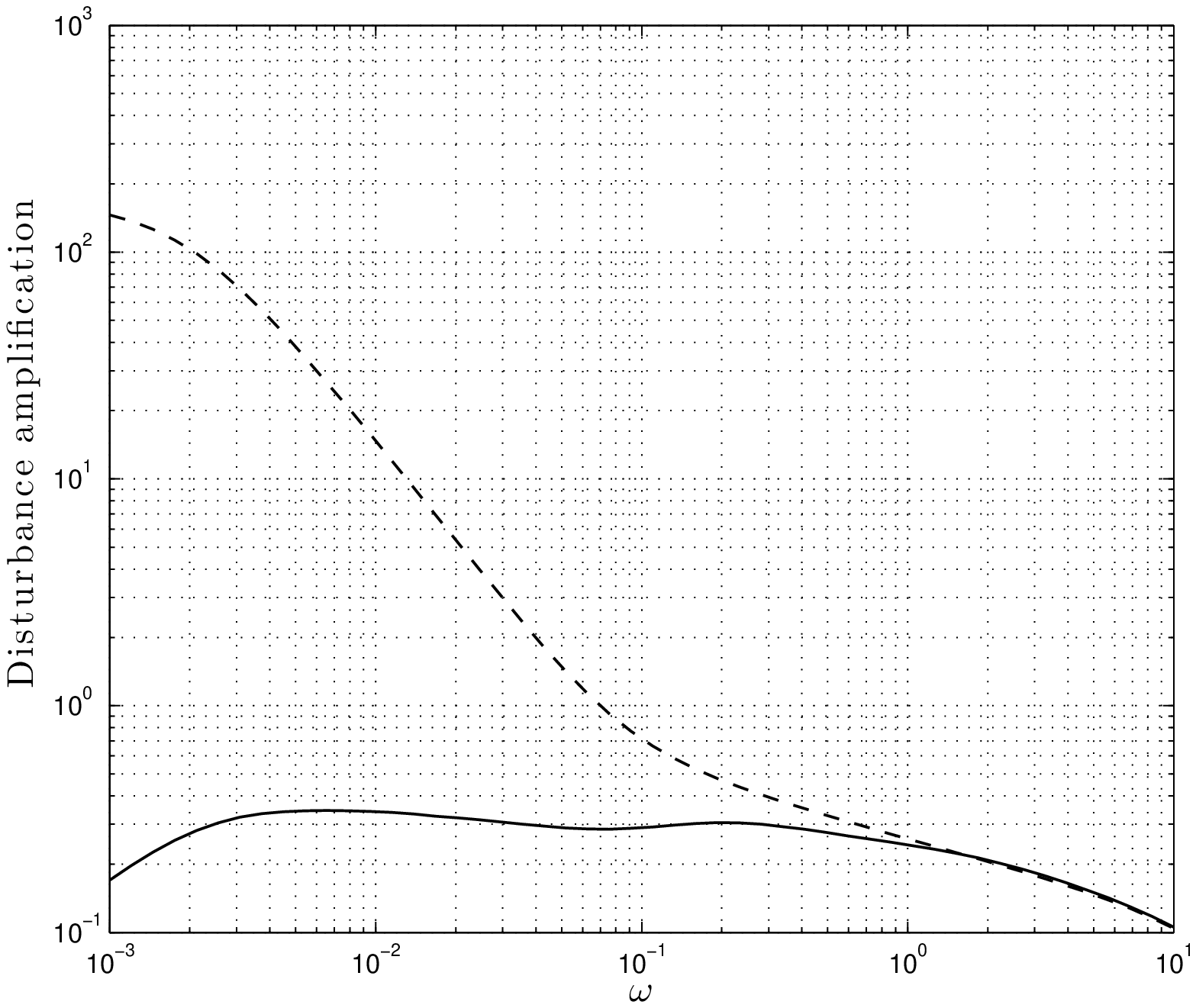}}
  \caption{Open (- -) and closed-loop~(--) disturbance responses, showing the range of temporal frequencies over which the loop-shaping controller attenuates the worst-case disturbance forcing~$\boldsymbol{f}$, arising form the nonlinearity of the flow, upon the wall-shear stress output~$\mathsfbi{y}$. Disturbance amplification is plotted in terms of~$\bar{\sigma}\left(\mathcal{P}_f(i\omega)\right)$ and~$\bar{\sigma}\left(\mathcal{P}_{\mathsfbi{y}f}(i\omega)\right)$, the respective singular value plots of the open and closed-loop transfer function matrices~$\mathcal{P}_f$ and~$\mathcal{P}_{\mathsfbi{y}f}$, defined in~\eqref{Pyf}.}
\label{Fig18}
\end{figure}

\begin{figure}
  \centerline{\includegraphics[scale=0.5]{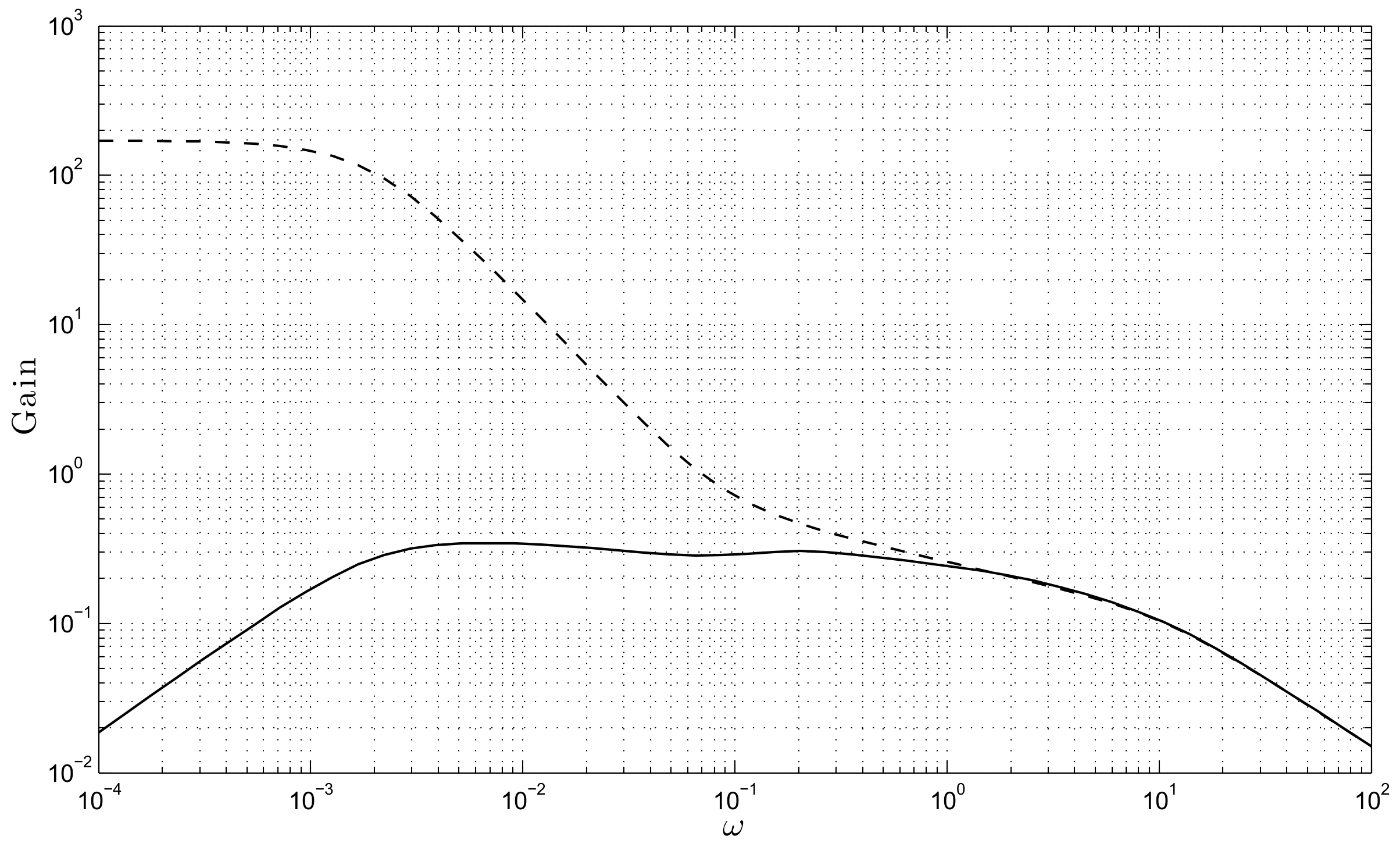}}
  \caption{Single-sided amplitude spectrum of the streamwise wall-shear stress perturbations for~$\Rey_\tau=210$. The magnitudes of the wall-shear stress perturbations are significantly lower for the controlled case~(--) at frequencies below the loop crossover frequency~$(\omega=0.3)$, compared to the uncontrolled flow~(- -). This is consistent with the linear system responses shown in~Figure~\ref{Fig18}.}
\label{Fig19}
\end{figure}

Further insight can be gained by studying the effects of the controller in the frequency domain. With reference to~\eqref{Pyf},~Figure~\ref{Fig18} displays the gain vs frequency plots for the open and closed-loop disturbance responses,~$\mathcal{P}_f$ and~$\mathcal{P}_{\mathsfbi{y}f}$, respectively, where~$\mathcal{P}_f$ shares the same dynamic model as~\eqref{channeldesc}, with the exception of homogenous boundary conditions in~$\mathsfbi{A}_\mathrm{D}$ and a disturbance input matrix~$\mathsfbi{B}_\mathrm{D}$ of the following form:
\begin{equation*}
	\mathsfbi{B}_\mathrm{D}:=\begin{bmatrix}\mathsfbi{H}^{-1}\\ 0\end{bmatrix},
\end{equation*}
where~$\mathsfbi{H}$ is a Cholesky factor of a discretised perturbation energy matrix~(see e.g.~\citet{Schmid}).~Figure~\ref{Fig18} shows that the controller provides attenuation of the input disturbances~$\boldsymbol{f}$, arising from the nonlinear forcing of the flow, across all frequencies. Compared to the uncontrolled flow, the controller attenuates disturbances up to a frequency of~$\omega=1$, just above the designed unity gain crossover frequency of the compensated system ($\omega_c=0.3$ in~Figure~\ref{Fig8}(b)). Again, and with reference to~\eqref{Pyfcl}, this is to be expected since the design of high loop-gain~$(\bar{\sigma}(\mathcal{PW}(i\omega))\gg 1)$ ensures that the effect of disturbances upon the system outputs are attenuated over a frequency range extending up to the vicinity of the unity-gain crossover frequency. Inspection of the spectral content of the wall-shear stress signals confirms this to be the case.~Figure~\ref{Fig19} displays the single-sided amplitude spectrum of the~DNS perturbation wall-shear stresses, from which it is apparent that the controller provides significant attenuation of these perturbations up to a frequency of around~$\omega=1$, after which the control action dies away. The disturbance attenuation of the controller is particularly noticeable at frequencies  below the crossover frequency. This is entirely consistent with the design of the precompensator in~Section~\ref{contdes}, where the use of integral control action provides high loop-gain, and hence high disturbance rejection, at low frequencies. The attenuation of low frequency disturbances is relevant since the forcing arising from the nonlinearity of the flow evolves over a longer timescale than the shear interaction timescale~\citep{landahl77:pof}, and hence contains lower frequencies. This explains why a linear controller, such as the one employed in this study, is able to attenuate the effects of the nonlinear forcing upon the wall-shear stress.

The same controller was then applied to flows perturbed across a range of~Reynolds numbers. At each Reynolds number, the perturbation RMS shear-stress was computed for the controlled and uncontrolled cases, and the percentage reductions are shown in~Table~\ref{Tab1}. Computational limitations limited the~DNS simulations to~$\Rey_\tau=360$ and under, but from~Table~\ref{Tab1} it is clear that the controller is providing effective attenuation of the wall-shear stress perturbations across a wide range of~Reynolds numbers. The controller is robust, not only to the parametric uncertainty induced by perturbing the Reynolds number of the flow, but also to the unmodelled dynamics arising form the use of a low-order spatially discretised model. This is to be expected following on from the results of the~$\upnu$-gap analysis in~Figure~\ref{Fig14}. In addition, the controller demonstrates robustness to the dynamic uncertainty arising from the nonlinearity of the flow, providing effective regulation of the wall-shear stress despite a turbulent initial condition in which the flow is significantly perturbed away from the laminar state assumed in the control model.
\begin{table}
\begin{center}
\begin{tabular}{c c c c c c c c}
$\Rey_\tau=175$ &  $\Rey_\tau=210$ &  $\Rey_\tau=247$ &  $\Rey_\tau=281$ &  $\Rey_\tau=315$ &  $\Rey_\tau=360$ \\ [3pt]
 87.9\% & 89.3\% & 87.5\% & 87.0\% & 88.1\% & 87.9\% \\
\end{tabular}
\caption{Percentage RMS reductions in perturbation wall-shear stresses under the control of the~$\mathcal{H}_\infty$ loop-shaping controller synthesised from the nominal flow model~$\mathcal{P}_{15,\mathcal{W}}^{5000}$.}\label{Tab1}
\end{center}
\end{table}

\section{Conclusions}\label{Sec8}
We have addressed the problem of obtaining models of systems based on the Navier-Stokes equations that provide a priori robust stability and performance bounds for closed-loop flow control. It is suggested that, from the point of view of employing existing linear control systems theory, there are essentially three problems to be tackled: linearisation, spatial discretisation and conversion from a system of DAEs to one of ODEs. We have presented results that add further evidence to suggest that linear control is effective in the control of wall turbulence even though turbulence is intrinsically nonlinear. Reasons for this are encapsulated in theories such as RDT, Landahl's ideas on sheared turbulence or gain-based analyses of turbulence formation. Specifically, by modelling the forcing arising from the nonlinearity of the flow as a disturbance input to the linear flow dynamics, we showed how the effects of such forcing could be heavily attenuated by designing a feedback controller with high loop-gain over a certain frequency range, and justified this range in terms of the timescale separation between linear and nonlinear mechanisms.

The present paper applied two methods for addressing the further issues of discretisation and conversion of equations from physical to state space.  For the first, the model-refinement procedure was applied to efficiently obtain spatially discretised models of low state dimension, from which robust controllers could be readily synthesised, with guaranteed performance bounds when applied to the actual flow. This is the first instance of this technique being applied to a flow control problem. Model refinement is the conceptual opposite of model-reduction based methods, since the starting point of the former approach lay with models of low, rather than high order, and where the emphasis lay upon obtaining models suitable for closed-loop, as opposed to open-loop control. This new approach to flow control employed established tools from robust control theory, such as the~$\upnu$-gap metric and the robust stability margin. In addition, it was argued that coprime factor uncertainty represents an appropriate choice of uncertainty model for capturing the inevitable discrepancies that exist between an actual fluid-flow system and a simpler control model, hence motivating the use of $\mathcal{H}_\infty$ loop-shaping control, a technique that to the best of our knowledge has not previously been applied to the problem of controlling wall turbulence. The problem of converting from physical to state space was overcome using a numerical approach that eased the prescription of boundary conditions, compared to traditional velocity vorticity-based methods.

These techniques were demonstrated upon a simplified problem of skin-friction drag reduction where the controller was based on a linearised, low-order model of the flow. A robust controller was synthesised that provided high attenuation of the perturbation streamwise wall-shear stresses across a wide range of Reynolds numbers, as evidenced by results from high fidelity linear simulations and nonlinear DNS. Actuation in the form of wall-based blowing/suction,~$v_{\mathrm{wall}}$, was based on sensing in~$\tau_{yx}$ such that perturbations of $\tau_{yx}$ for a given wavenumber pair were minimised. In doing so, the control exerted an influence over~$\tau_{yz}$ by setting up near-wall `buffer' vortices.

Future research should address the problem of optimal actuator and sensor placement. Importantly, the framework of gap-metrics and~$\mathcal{H}_\infty$ loop-shaping can be extended to address this very issue, as demonstrated by~\citet{RS03}. Loop shaping design is a (temporal) frequency domain approach to control systems design. However, it should be noted that frequency does not unambiguously distinguish large structure moving quickly from small structure moving slowly, and it is the former that makes the greater contribution to skin friction. In the present work, we have assumed, for simplicity, that the walls are densely populated with sensors. As a result, the linearised system is rendered observable.  Similarly, if in the control objective, small structure plays an important part in generating skin-friction drag then the techniques described in this paper ensures mesh refinement to an appropriate level by increasing the spatial fidelity of the model. Hence, the model refinement process indicates the appropriate degree of refinement required to meet the control objective. For more realistic configurations, future research should also address the consequences of non-conservative domains, i.e. those in which the nonlinearity of the disturbance field may be taken to be significant, and the extent to which it may be accommodated by the disturbance rejection framework. For flows exhibiting a more broadband forcing, such as in a turbulent mixing layer for example, our approach would require a high bandwidth controller to reject high frequency disturbances, which would likely necessitate the use of fast actuation, which may or may not be possible. The leads onto the final point that despite the potential effectiveness of the linear controllers developed in this paper, it is possible that some form of nonlinear control may provide enhanced performance by selectively exploiting the nonlinearity of the flow in some desirable fashion, and designing such controllers could be an interesting avenue of future research.


\begin{appendix}
\section{Quantifying the unknown with coprime factor uncertainty}\label{CFU}
\begin{figure}
\centerline{\includegraphics[scale=0.07]{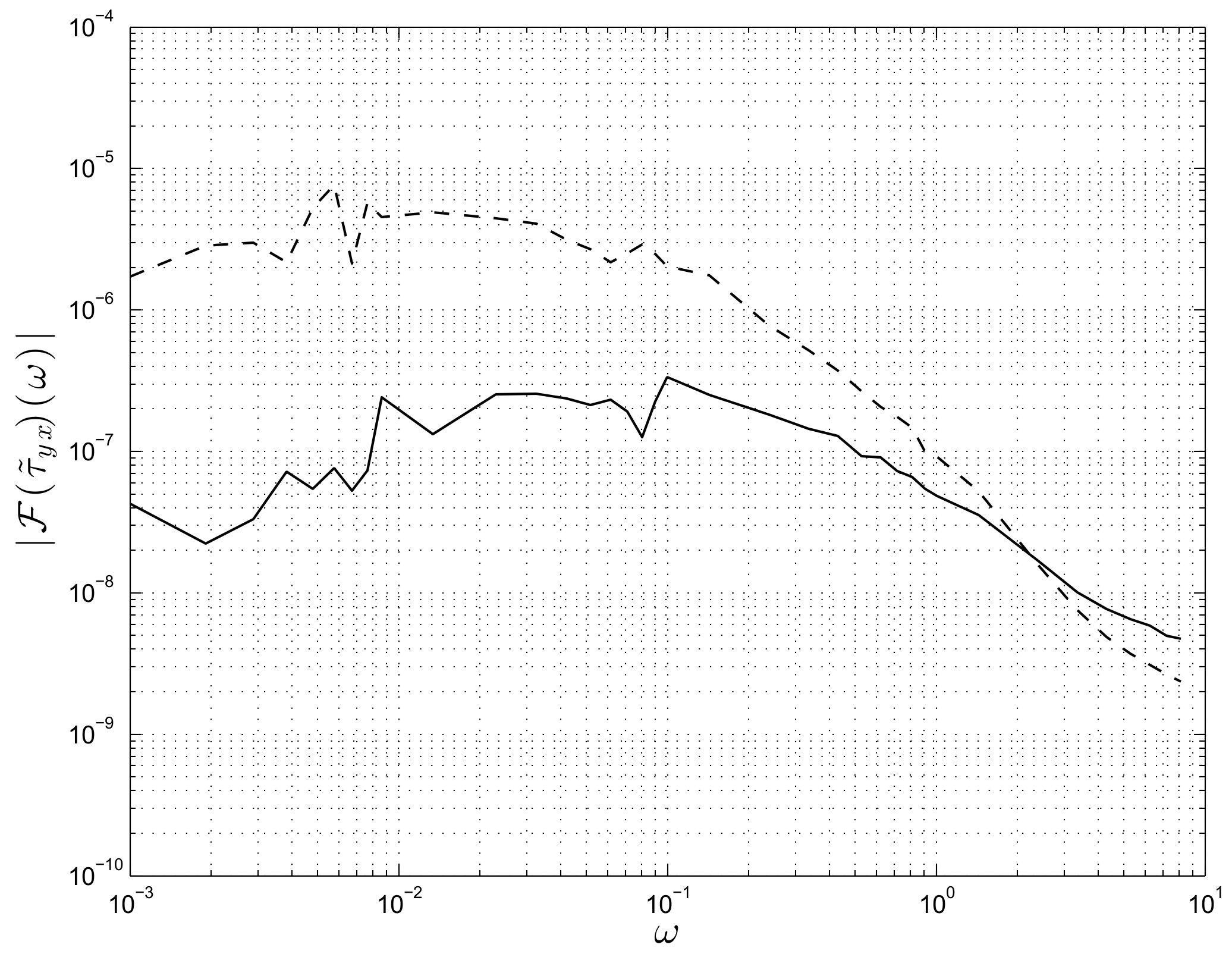}}
  \caption{Coprime factor uncertainty. The shaded box represents the perturbed plant, and~$\mathsfbi{v}_1$,~$\mathsfbi{v}_2$,~$\mathsfbi{w}_1$,~$\mathsfbi{w}_2$ represent disturbances entering the system at different points.}
\label{Fig20}
\end{figure}

Coprime factor perturbations take the form:
\begin{equation}
	\mathcal{P}_{\textrm{p}}:=\left\{(\mathcal{N}+\mathcal{U}_\mathcal{N})(\mathcal{M}+\mathcal{U}_\mathcal{M})^{-1}\right\},~~\textrm{such that~~}\norm{\begin{bmatrix}\mathcal{U}_\mathcal{N}\\ \mathcal{U}_\mathcal{M}\end{bmatrix}}_{\infty}<\frac{1}{\gamma},
\end{equation}
with~$\gamma>1$ and where~$\mathcal{P}=\mathcal{NM}^{-1}$ is a normalised right coprime factorisation of the unperturbed plant model~$\mathcal{P}$, meaning $\mathcal{M}^*\mathcal{M}+\mathcal{N}^*\mathcal{N}=\mathsfbi{I}$. The relevant block diagram is depicted in~Figure~\ref{Fig20} where the signals~$\mathsfbi{v}_1$ and~$\mathsfbi{v}_2$ represent disturbances on the control inputs~$\mathsfbi{u}$ and measurements~$\mathsfbi{y}$, respectively, whilst~$\mathsfbi{w}_1$ and~$\mathsfbi{w}_2$ represent disturbances acting upon the plant. The transfer functions relating~$\mathsfbi{u}$ and~$\mathsfbi{y}$, to the disturbances, are:
\begin{equation}\label{transfncs}
	\begin{bmatrix}\mathsfbi{y}\\ \mathsfbi{u}\end{bmatrix}=
	\begin{bmatrix}\mathcal{P}\\ \mathsfbi{I} \end{bmatrix}\begin{matrix}(\mathsfbi{I}-\mathcal{KP})^{-1}\\ ~ \end{matrix}\begin{matrix} \begin{bmatrix}-\mathcal{K} & \mathsfbi{I} \end{bmatrix}\\ ~ \end{matrix}\begin{bmatrix}\mathsfbi{v}_2\\ \mathsfbi{v}_1\end{bmatrix}+\begin{bmatrix}\mathsfbi{I}\\ \mathsfbi{K} \end{bmatrix}\begin{matrix}(\mathsfbi{I}-\mathcal{PK})^{-1}\\ ~ \end{matrix}\begin{matrix} \begin{bmatrix}\mathsfbi{I} & -\mathcal{P} \end{bmatrix}\\ ~ \end{matrix}\begin{bmatrix}\mathsfbi{w}_2\\ \mathsfbi{w}_1\end{bmatrix},
\end{equation}

It can be shown~\citep{UncandFeed} that it is precisely the norm of the first of these transfer function matrices that should be minimised in order to obtain robust stability with respect to perturbations to the normalised coprime factors of the model. This motivates the following definition:
\newtheorem{defn}{Definition}[section]
\begin{defn} The stability margin for coprime factor perturbations $b_{\mathcal{P},\mathcal{K}}$ is defined as follows~\citep{UncandFeed}:
\begin{equation}
	b_{\mathcal{P},\mathcal{K}}:=\left\{ \begin{array}{ll}
	\norm{\begin{bmatrix}\mathcal{P}\\ \mathsfbi{I} \end{bmatrix}\begin{matrix}(\mathsfbi{I}-\mathcal{KP})^{-1}\\ ~ \end{matrix}\begin{matrix} \begin{bmatrix}-\mathcal{K} & \mathsfbi{I} \end{bmatrix}\\ ~ \end{matrix}}^{-1}_{\infty}
	& \textrm{if $[\mathcal{P}, ~\mathcal{K}]$ is stable,}\\
0 & \textrm{otherwise,}
\end{array} \right.
\end{equation}
\end{defn}

If follows that  $b_{\mathcal{P},\mathcal{K}}\geq \frac{1}{\gamma}$ and so a natural objective is to make $b_{\mathcal{P},\mathcal{K}}$ as large as possible, subject to design criteria~\citep{McFarlane}. By way of illustration, for single input single output (SISO) systems it can be shown~\citep{UncandFeed} that a feedback system with~$b_{\mathcal{P},\mathcal{K}}=0.3$ provides reasonable gain and phase margins~\citep{Astrom} of~\emph{at least}~$2$ and~$35^{\circ}$, respectively. Hence,~$b_{\mathcal{P},\mathcal{K}}$ can be thought of as a generalisation of classical gain and phase margins to systems with multiple inputs and multiple outputs.

It can also be shown~\citep{UncandFeed} that the $\mathcal{H}_{\infty}$ norm of the second transfer function in~\eqref{transfncs} is equal to that of the first, i.e.
\begin{equation}
	\norm{\begin{bmatrix}\mathcal{P}\\ \mathsfbi{I} \end{bmatrix}\begin{matrix}(\mathsfbi{I}-\mathcal{KP})^{-1}\\ ~ \end{matrix}\begin{matrix} \begin{bmatrix}-\mathcal{K} & \mathsfbi{I} \end{bmatrix}\\ ~ \end{matrix}}_{\infty} =
	\norm{\begin{bmatrix}\mathsfbi{I}\\ \mathcal{K} \end{bmatrix}\begin{matrix}(\mathsfbi{I}-\mathcal{PK})^{-1}\\ ~ \end{matrix}\begin{matrix} \begin{bmatrix}\mathsfbi{I} & -\mathcal{P} \end{bmatrix}\\ ~ \end{matrix}}_{\infty}.
\end{equation}

Consequently the use of $b_{\mathcal{P},\mathcal{K}}$ as a measure of robust stability as well as robust performance is easily motivated by noting that it bounds the gain of all eight closed-loop transfer functions, between inputs and outputs at any point in the loop.

\section{Computing a standard state-space system from descriptor form}\label{computesf}
Let~$\mathsfbi{E}_\mathrm{D},~\mathsfbi{A}_\mathrm{D}\in\mathbb{C}^{n_\mathrm{D}\times n_\mathrm{D}}$. The pair~$(\mathsfbi{E}_\mathrm{D},\mathsfbi{A}_\mathrm{D})$ is said to be~\emph{regular} if there exists an $s\in\mathbb{C}$ such that $\mathrm{det}(s\mathsfbi{E}_\mathrm{D}-\mathsfbi{A}_\mathrm{D})\neq 0$~\citep{Dai}. If~$(\mathsfbi{E}_\mathrm{D},\mathsfbi{A}_\mathrm{D})$ in~\eqref{descriptora} is regular, there exist nonsingular matrices $\mathsfbi{T},\mathsfbi{S}\in\mathbb{C}^{r\times r}$ such that the transformation:
\begin{subequations}\label{standardform}
\begin{equation}\label{PandS}
	\mathsfbi{T}\mathsfbi{E}_\mathrm{D}\mathsfbi{S}\mathsfbi{S}^{-1}\dot{\mathsfbi{x}}_\mathrm{D}(t)=\mathsfbi{T}\mathsfbi{A}_\mathrm{D}\mathsfbi{S}\mathsfbi{S}^{-1}\mathsfbi{x}_\mathsfbi{D}(t)+\mathsfbi{T}\mathsfbi{B}_\mathrm{D}\mathsfbi{u}(t),
\end{equation}
yields the following system in standard form~\citet[Lem. 2.3]{Gerdin}:
\begin{align}\label{standardformb}
	\begin{bmatrix} \mathsfbi{I} & 0\\ 0 & \mathsfbi{N}\end{bmatrix}\begin{bmatrix}\dot{\mathsfbi{x}}(t)\\ \dot{\chi}(t)\end{bmatrix}=
	\begin{bmatrix} \mathsfbi{A} & 0\\ 0 & \mathsfbi{I}\end{bmatrix}\begin{bmatrix}\mathsfbi{x}(t)\\ \chi(t)\end{bmatrix}+\begin{bmatrix}\mathsfbi{B}\\ \mathsfbi{G}\end{bmatrix}\mathsfbi{u}(t),
\end{align}
\end{subequations}
where~$\mathsfbi{N}\in\mathbb{C}^{(n_\mathrm{D}-n)\times(n_\mathrm{D}-n)}$ is nilpotent (meaning that~$\mathsfbi{N}^{i_{\textrm{np}}}=0$ for some~$i_{\textrm{np}}\in\mathbb{N}$),~$\mathsfbi{A}$ and~$\mathsfbi{B}$ are as in~\eqref{standard},~$\mathsfbi{G}\in\mathbb{C}^{(n_\mathrm{D}-n)\times m}$,~$\mathsfbi{I}$ are identity matrices of compatible dimensions and~$\left[\begin{smallmatrix}\mathsfbi{x}(t)\\\chi(t)\end{smallmatrix}\right]=\mathsfbi{S}^{-1}\mathsfbi{x}_\mathrm{D}(t)$. 
The matrices in~\eqref{standardform} are computed as follows~\citep{Gerdin, Schon,Shahzad11}:
\begin{enumerate}
\renewcommand{\labelenumi}{(\roman{enumi})}
\item~Compute the generalised Schur form of the matrix pencil~$\lambda \mathsfbi{E}_\mathrm{D} -\mathsfbi{A}_\mathrm{D}$ so that:
	\begin{align}\label{qz}
	\mathsfbi{T}_1(\lambda \mathsfbi{E}_\mathrm{D}-\mathsfbi{A}_\mathrm{D})\mathsfbi{S}_1 = \lambda\left[\begin{smallmatrix}\mathsfbi{E}_1 & \mathsfbi{E}_2\\ 0 & \mathsfbi{E}_3\end{smallmatrix}\right]
	+\left[\begin{smallmatrix}\mathsfbi{A}_1 & \mathsfbi{A}_2\\ 0 & \mathsfbi{A}_3\end{smallmatrix}\right],
	\end{align}
where~$\mathsfbi{T}_1$ and~$\mathsfbi{S}_1$ are unitary matrices i.e.~$\mathsfbi{T}_1^*\mathsfbi{T}_1=\mathsfbi{T}_1\mathsfbi{T}_1^*=\mathsfbi{I}$, and are not to be confused with~$\mathsfbi{T}$ and~$\mathsfbi{S}$ in~\eqref{PandS}. The generalised eigenvalues should be sorted so that the diagonal elements of~$\mathsfbi{E}_1$ contain only non-zero elements. The generalised Schur form and the subsequent reordering can be computed using a QZ algorithm~\citep{Golub}.
\item~Solve the following coupled Sylvester equation to obtain the matrices~$\mathsfbi{L}$ and~$\mathsfbi{R}$:
	\begin{subequations}\label{gensylv}
	\begin{align}
	\mathsfbi{E}_1 \mathsfbi{R} + \mathsfbi{L}\mathsfbi{E}_3 & = -\mathsfbi{E}_2,\label{gensylva} \\
	\mathsfbi{A}_1 \mathsfbi{R} + \mathsfbi{L}\mathsfbi{A}_3 & = -\mathsfbi{A}_2.\label{gensylvb}
	\end{align}
	\end{subequations}
The solution to~\eqref{gensylv} can be obtained by solving for~$\mathsfbi{L}$ in:
\begin{subequations}\label{mysylv}
\begin{equation}\label{mysylv1}
\mathsfbi{A}_1 \mathsfbi{E}_1^{-1} \mathsfbi{L} \mathsfbi{E}_3 \mathsfbi{A}_3^{-1} -\mathsfbi{L} - \left(\mathsfbi{A}_2-\mathsfbi{A}_1 \mathsfbi{E}_1^{-1}\mathsfbi{E}_2\right)\mathsfbi{A}_3^{-1}=0,
\end{equation}
and substituting to obtain~$\mathsfbi{R}$:
\begin{equation}\label{mysylv2}
\mathsfbi{R}=-\mathsfbi{E}_1^{-1}\mathsfbi{E}_2-\mathsfbi{E}_1^{-1}\mathsfbi{L}\mathsfbi{E}_3.
\end{equation}
\end{subequations}
\item~Form the matrices in~\eqref{standardform} as follows:
	\begin{subequations}\label{sfmatrices}
	\begin{gather}
	\mathsfbi{T}=\begin{bmatrix}\mathsfbi{E}_1^{-1} & 0\\ 0 & \mathsfbi{A}_3^{-1}\end{bmatrix}\begin{bmatrix}\mathsfbi{I} & \mathsfbi{L}\\ 0 & \mathsfbi{I}\end{bmatrix}\mathsfbi{T}_1,~~
	\mathsfbi{S}=\mathsfbi{S}_1\begin{bmatrix}\mathsfbi{I} & \mathsfbi{R}\\ 0 & \mathsfbi{I}\end{bmatrix},\label{sfmatricesa}\\
	\mathsfbi{A}=\mathsfbi{E}_1^{-1}\mathsfbi{A}_1,\quad
	\begin{bmatrix}\mathsfbi{B}\\ \mathsfbi{G}\end{bmatrix}=\mathsfbi{T}\mathsfbi{B}_\mathrm{D},\quad
	\mathsfbi{N}=\mathsfbi{A}_3^{-1}\mathsfbi{E}_3.\label{sfmatricesb}
	\end{gather}
	\end{subequations}
\item~Provided~\citep{Gerdin}~$\mathsfbi{N}^j \mathsfbi{G}=0$ for all~$j\in\mathbb{N}\geq 1$, then~\eqref{descriptorb} is given by:
\begin{align}\label{sfoutput}
	\mathsfbi{y}(t)=\mathsfbi{C}_\mathrm{D}\mathsfbi{S}\begin{bmatrix}\mathsfbi{x}(t)\\\chi(t)\end{bmatrix}&=\begin{bmatrix}\mathsfbi{C} & \mathsfbi{J}\end{bmatrix}\begin{bmatrix}\mathsfbi{x}(t)\\\chi(t)\end{bmatrix}\nonumber\\
	&~~~~~=\mathsfbi{C}\mathsfbi{x}(t)+\mathsfbi{D}\mathsfbi{u}(t).
\end{align}
where~$\mathsfbi{J}\in\mathbb{C}^{q\times(n_\mathrm{D}-n)}$ and $\mathsfbi{D}:=-\mathsfbi{J}\mathsfbi{G}$. This completes the numerical conversion of a descriptor state-space system~\eqref{descriptor} into a standard state-space system~\eqref{standard}.
\end{enumerate}

\end{appendix}

\bibliographystyle{jfm}
\bibliography{biblio}

\end{document}